\documentclass[aps,prb,twocolumn,floatfix]{revtex4-1}

\pdfoutput=1

\usepackage{times}
\usepackage{amsfonts}
\usepackage{amssymb}
\usepackage{amsmath}
\usepackage{graphicx}
\usepackage{bm}
\usepackage{verbatim}

\usepackage{bm}
\usepackage{color}
\usepackage{stackrel}
\usepackage{accents}
\usepackage{hyperref}
\usepackage{latexsym}

\usepackage{ulem}

\newcommand{\kk}{\mathbf{k}}
\newcommand{\Aq}{\mathbf{A}}



\begin{document}

\title{Collective modes in pumped unconventional superconductors with competing ground states}

\author{Marvin A. M\"{u}ller$^{1}$, Pavel A. Volkov$^{1,2}$, Indranil Paul$^3$,  and Ilya M. Eremin$^1$}
\affiliation{$^1$ Institut f\"{u}r Theoretische Physik III, Ruhr-Universit\"{a}t Bochum, D-44801 Bochum, Germany \\
$^2$Department of Physics and Astronomy, Center for Materials Theory, Rutgers University, Piscataway, NJ 08854 USA \\
$^3$Laboratoire Mat\'{e}riaux et Ph\'{e}nom\`{e}nes Quantiques, Universit\'{e} de Paris, CNRS, F-75013, Paris, France}

\begin{abstract}
Motivated by the recent development of terahertz pump-probe experiments, we investigate the short-time dynamics in superconductors with multiple attractive pairing channels. Studying a single-band square lattice model with spin-spin interaction as an example, we find the signatures of collective excitations of the  pairing symmetries (known as Bardasis-Schrieffer modes) as well as the order parameter amplitude (Higgs mode) in the short-time dynamics of the spectral gap and quasiparticle distribution after an excitation by a pump pulse. We show that the polarization and intensity of the pulse can be used to control the symmetry of the non-equilibrium state as well as frequencies and relative intensities of the contributions of different collective modes. We find particularly strong signatures of the Bardasis-Schrieffer mode in the dynamics of the quasiparticle distribution function. Our work shows the potential of modern ultrafast experiments to address the collective excitations in unconventional superconductors and highlights the importance of sub-dominant interactions for the non-equilibrium dynamics in these systems.
\end{abstract}

\maketitle

%\section{Introduction}

Ultrafast pump-probe techniques became recently a powerful tool to probe the temporal evolution of symmetry broken states and relaxation in conventional and unconventional superconductors.\cite{Pashkin2010,Beck2011,Cav2011Science,Matsunaga2012,Conte2012,Beck2013,Mansart2013,Shimano2014Science,CavNature2014,CavPRB2015,Matsunaga2017,Katsumi2018} An intense pulse
couples non-linearly to the Cooper pairs  of the superconductor and, as was argued theoretically, should lead to a coherent excitation of the
Higgs amplitude mode, i.e.  $|\Delta(t)|$ performs a damped oscillation with frequency $\omega_H = 2|\Delta(\infty)|$.\cite{Volkov1974,Amin2004,Barankov2004,Yuzbashyan2005,Yuzbashyan2006,Barankov2006,Papenkort2007,Krull2014,Big-Quench-Review2015,Tsuji2015,Aoki-Higgs2,Matt-Higgs} Nonlinear terahertz spectroscopy has enabled the observation of the Higgs mode
in conventional superconductors in the form of a free or forced oscillation and the resulting third-harmonic generation\cite{Matsunaga2012,Matsunaga2014,Matsunaga2017}. Interestingly, this technique has been also
recently applied to the unconventional superconductors such as high-T$_c$ cuprates with the $d$-wave symmetry of the superconducting gap \cite{Katsumi2018,Chu2019} where some additional oscillations have been reported.\cite{Chu2019}

In contrast to conventional superconductors, where the pairing is driven by the attractive electron-phonon interaction, the pairing interaction in unconventional superconductors is most likely of repulsive nature. To overcome the net repulsion among the quasiparticles, the superconducting gap has to change its sign across different parts of the Fermi surface, which typically yields the superconducting gap of a lower symmetry than an isotropic $s$-wave.
For example, it is generally known that the antiferromagnetic spin fluctuations peaked near wave vector ${\bf Q}_{AF}=(\pi,\pi)$ within a single-band model on a square lattice give rise to a $d_{x^2-y^2}$-wave symmetry of the superconducting gap, yet states having other symmetries, such as strongly anisotropic sign-changing (extended) $s$-wave symmetry and the $d_{xy}$-symmetry, are closely competing.
As a result, the temporal dynamics of single-band unconventional superconductors might be significantly richer than that of the conventional ones.\cite{Barankov2004,Papenkort2007,Yuzbashyan2005,Tsuji2015}

In this manuscript we analyze the short-time dynamics in a single-band unconventional superconductor with multiple competing pairing symmetries. In particular, we consider a single-band model of fermions on a square lattice interacting via spin-spin interaction. The interaction can be decoupled into various pairing channels with different symmetry. Varying the band filling we find two competing even-parity superconducting states forming a typical phase diagram of an unconventional superconductor where different ground states can be accessed by doping. Studying the system driven out of equilibrium by a laser pulse, we show how the collective signatures of symmetries different from a given ground state symmetry, known as Bardasis-Schrieffer modes\cite{bardasis61,maiti15,maiti16,Boehm18} in the context of $s$-wave ground state in the equilibrium, evolve as a function of doping and the polarization direction. Depending on the polarization direction of the incoming light the tetragonal symmetry is broken, which necessarily leads to a mixing of $s$- and $d$-wave symmetries in non-equilibrium. Furthermore, we show that the particle distribution acquires an additional $d_{x^2-y^2}$-character and oscillates dominantly with the Bardasis-Schrieffer mode frequency, which may be observed in time-resolved ARPES experiments.
Our study highlights the important role of sub-dominant pairing states in the short-time dynamics of unconventional superconductors and we identify the signatures of the resulting collective modes that can be observed in future pump-probe experiments.

\begin{figure}	
	\includegraphics[width = 0.8\linewidth]{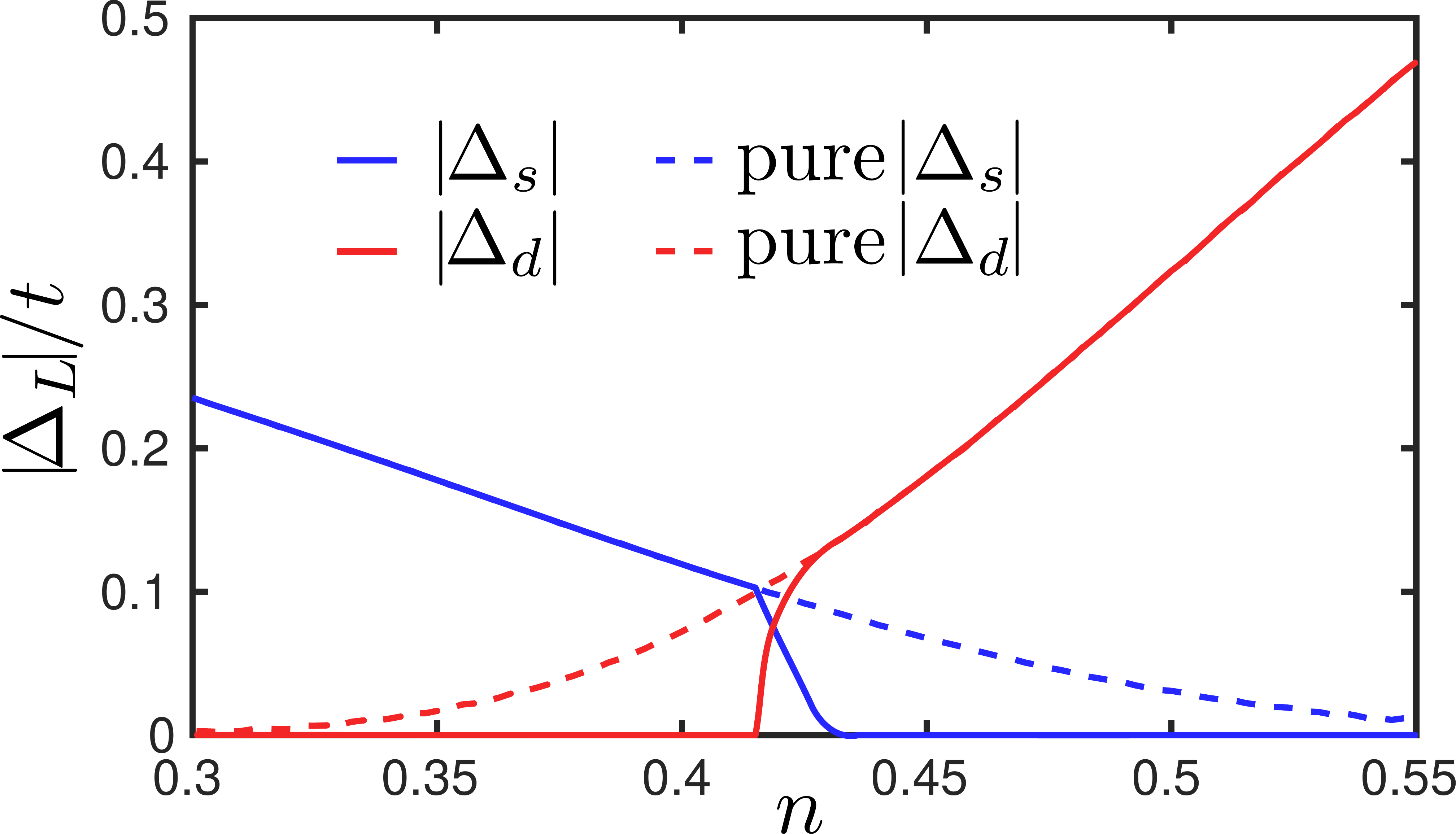}
	\caption{Phase diagram of $H_{\text{MF}}$ at $T=0$ for the fillings where $d_{x^2-y^2}$-wave order parameter closely competes with that of the extended $s$-wave symmetry. The solid lines refer to the actual ground state values for $\Delta_s$ and $\Delta_{d_{x^2-y^2}}$, while the dashed lines show the behavior of a pure $s$-wave or pure $d_{x^2-y^2}$-wave solution, ignoring the possibility of coexistence.
}
\label{fig:phasediagram}
\end{figure}

Our starting point is a model of fermions interacting via a spin-spin interaction
\begin{equation}
	H =-t\sum_{\langle ij\rangle,\sigma} c_{i \sigma}^\dagger c_{j \sigma} + J \sum_{\langle ij \rangle} \mathbf{S}_i\cdot\mathbf{S}_j,
\end{equation}
where $c_{i,\sigma}^{(\dagger)}$ are fermionic annihilation (creation) operators on site $i$ and spin $\sigma$, $t$ is the hopping integral between nearest neighbors and $S_i^{\alpha} = \frac{1}{2} \sum_{s,s^{\prime}} c_{i s}^\dagger \sigma_{ss^{\prime}}^{\alpha}c_{i s^{\prime}}$ are the spin-1/2 operators.
This model was considered before in the context of cuprates, pnictides and heavy fermion systems.\cite{Davis2013} In the momentum space the tight-binding energy dispersion is given by ${\xi_\kk = -2t(\cos(k_x) + \cos(k_y)) -\mu}$. The spin-spin interaction can also be transformed into momentum space and decoupled into a number of superconducting channels
\begin{align}
	J_{\kk,\kk^\prime} \equiv & V_{s}\gamma_{\kk,s}\gamma_{\kk^\prime,s} + V_{d_{x^2-y^2}}\gamma_{\kk,d_{x^2-y^2}}\gamma_{\kk^\prime,d_{x^2-y^2}} \nonumber \\  &+ V_{p_x}\gamma_{\kk,p_x}\gamma_{\kk^\prime,p_x} + V_{p_y}\gamma_{\kk,p_y}\gamma_{\kk^\prime,p_y},
\end{align}
where ${V_s = V_{d_{x^2-y^2}} = -3J/2}$ is the even-parity spin-singlet interaction with ${\gamma_{\kk,s} = (\cos(k_x) + \cos(k_y))/2}$ and ${\gamma_{\kk,d_{x^2-y^2}} = (\cos(k_x) - \cos(k_y))/2}$ form factors, respectively. The remnant components are repulsive odd-parity spin-triplet coupling constants ${V_{p_x} = V_{p_y} = J/2}$ with the $p_x$- and $p_y$-wave form factors ${\gamma_{\kk,p_{x(y)}} = \sin(k_{x(y)})}$. As we show below, the light pulse we consider does not couple the odd- and even- parity channels. As we focus on the regime, where the ground state has even parity, the odd and even parity solutions cannot mix both in- and out of equilibrium.
In the mean-field approximation the Hamiltonian reduces to
\begin{align}\label{eq:hamil}
    H_{\text{MF}} \simeq \sum_{\kk \sigma} \xi_{\kk}c_{\kk \sigma}^\dagger c_{\kk \sigma} + \sum_{\kk,l} \left[\Delta_l\gamma_{\kk,l} c_{\kk\uparrow}^\dagger c_{-\kk\downarrow}^\dagger + \text{h.c.} \right]
\end{align}
where $l=s,d_{x^2-y^2}$ and the ${\Delta_l = -V_l\sum_\kk \gamma_{\kk,l} \langle c_{-\kk\downarrow}c_{\kk\uparrow}\rangle}$ are the $s$- and $d_{x^2-y^2}$-wave component of the total superconducting order parameter $ {\Delta_\kk = \Delta_s\gamma_{\kk,s} + \Delta_{d_{x^2-y^2}}\gamma_{\kk,d_{x^2-y^2}}}$. 
In what follows, we work at $T=0$. Minimizing the energy, we obtain the equilibrium values of the superconducting order parameters $\Delta_s,\;\Delta_{d_{x^2-y^2}}$ as a function of $n$, shown in Fig. \ref{fig:phasediagram}. We concentrate on the region of the phase diagram where $s$-wave and $d$-wave symmetries are neighbors, but avoid the regime of coexistence of both order parameters, i.e. a possible $s+id$ state, as this state (and the similar case of $d+id$ state) and its non-equilibrium dynamics were discussed previously, \cite{maiti15,mueller18,Kirmani2018}. Instead, we focus on pairing fluctuations of the sub-dominant symmetry close to the transition points.
To simplify further calculations, we introduce the Anderson pseudospin notation\cite{Anderson1958}
\begin{align}
    \mathbf{s}_\kk = \frac{1}{2} \left( c_{\kk\uparrow}^\dagger,  c_{-\kk\downarrow}\right) \boldsymbol{\sigma} \begin{pmatrix} c_{\kk\uparrow} \\ c_{-\kk\downarrow}^\dagger\end{pmatrix},
\end{align}
where $\boldsymbol{\sigma} = \left(\sigma_x,\sigma_y,\sigma_z\right)^T$ are the Pauli-matrices. Using this vector, one can recast the Hamiltonian
%\begin{align}\label{eq:hamil_spin}
    $H = \sum_\kk \mathbf{B}_{\kk}\cdot \mathbf{s}_{\kk}$,
%\end{align}
which has the form of a set of (pseudo-)spins $\mathbf{s}_{\kk}$ coupled to a (pseudo-)magnetic field ${\mathbf{B}_{\kk} = \left(2\Delta^\prime_\kk,2\Delta^{\prime\prime}_\kk,2\xi_\kk\right)^T}$, with the notation $\Delta_\kk = \Delta_\kk^\prime -i \Delta_\kk ^{\prime\prime}$.
At zero temperature the thermal expectation values of the pseudospin components are given by ${\left\langle s_\kk^x \right\rangle = -\frac{\Delta^\prime_\kk}{2E_{\kk}}}$, $\left\langle s_\kk^y \right\rangle =  -\frac{\Delta^{\prime\prime}_\kk}{2E_{\kk}}$, and ${\left\langle s_\kk^z \right\rangle =   -\frac{\xi_\kk}{2E_{\kk}}}$
with the quasiparticle energy dispersion $E_\kk = \sqrt{\xi_\kk^2 + |\Delta_\kk|^2}$.

To investigate the collective modes in our model, we study the equations of motion for the pseudospin expectation values ${\mathbf{s}_\kk}$ that have the form of Bloch equations:
\begin{align}\label{eq:eom}
    \frac{d}{dt}\langle\mathbf{s}_\kk\rangle = \mathbf{B}_\kk \times \langle\mathbf{s}_\kk\rangle.
\end{align}
This equation together with the self-consistency equation for the superconducting gaps $\Delta_{s,d}(t) = -\sum_\kk V_{s,d}\gamma_{\kk,s,d}( \left\langle s_\kk^x \right\rangle(t) - i\left\langle s_\kk^y \right\rangle(t))$ yields a closed set of coupled differential equations, which defines the temporal evolution of all relevant quantities. To drive the system out of equilibrium we model the electric field of a laser pulse by including a time-dependent vector potential $\mathbf{A}(t)$ via the Peierls substitution. This results in ${\mathbf{B}_\kk = \left(2\Delta^\prime_\kk,2\Delta^{\prime\prime}_\kk,\xi_{\kk + \frac{e}{c}\mathbf{A}} + \xi_{\kk - \frac{e}{c}\mathbf{A}}\right)^T}$ (see Supplementary Material for details).   

To consider the possibility of exciting order parameter symmetries, different from the ground state one it is instructive to  split the equations of motion in Eq. \eqref{eq:eom} into different symmetry channels. As the pulse temporarily breaks $C_4$-rotational symmetry, different symmetry representations may mix. In particular, we can decompose the pseudomagnetic field into all even parity irreducible representation for the tetragonal $D_{4h}$ symmetry as follows: ${\mathbf{B}_\kk = \mathbf{B}_{\kk,s}  + \mathbf{B}_{\kk,d_{x^2-y^2}}  + \mathbf{B}_{\kk,d_{xy}} + \mathbf{B}_{\kk,g_{xy(x^2-y^2)}}}$, where we define $\mathbf{B}_{\kk,l} = \left(2\Delta_l^\prime \gamma_{\kk,l}, 2\Delta_l^{\prime\prime} \gamma_{\kk,l}, \xi_{\kk,\Aq,l}\right)^T$, where we have introduced the symmetrized notations for $B_{\kk}^z = \xi_{\kk + \frac{e}{c}\mathbf{A}} + \xi_{\kk - \frac{e}{c}\mathbf{A}} \equiv \xi_{\kk,\Aq,s} + \xi_{\kk,\Aq,d_{x^2-y^2}} + \xi_{\kk,\Aq,d_{xy}} + \xi_{\kk,\Aq,g_{xy(x^2-y^2)}}$. There appears no odd-parity component of the pseudomagnetic field $\mathbf{B_\kk}$, as can be seen immediately from the definition above.

 Let us now discuss the symmetry mixing in non-equilibrium. Denoting the ground state symmetry as $l_0$, one finds that   $\frac{d}{dt}\langle s_{\kk,l^\prime}^x\rangle=0$ and
$\frac{d}{dt}\langle \mathbf{s}_{\kk,l^\prime}^y\rangle = \mathbf{B}_{\kk,l}\times \langle \mathbf{s}_{\kk,l_0} \rangle $, where $l^\prime = l\otimes l_0$  is the symmetry of a product of functions with symmetries $l$ and $l_0$. Therefore, a finite $l$-symmetry component of $\mathbf{B}_\kk$ field is needed to induce pairing correlations of symmetry $l^\prime$. Moreover, the self-consistency equations imply that inducing an out-of equilibrium order parameter $\Delta_{l^\prime}$ additionally requires a non-zero $V_{l^\prime}$, making the consideration of the sub-dominant pairing interaction crucial.
\begin{figure}[h]	
	\includegraphics[width = 0.9\linewidth]{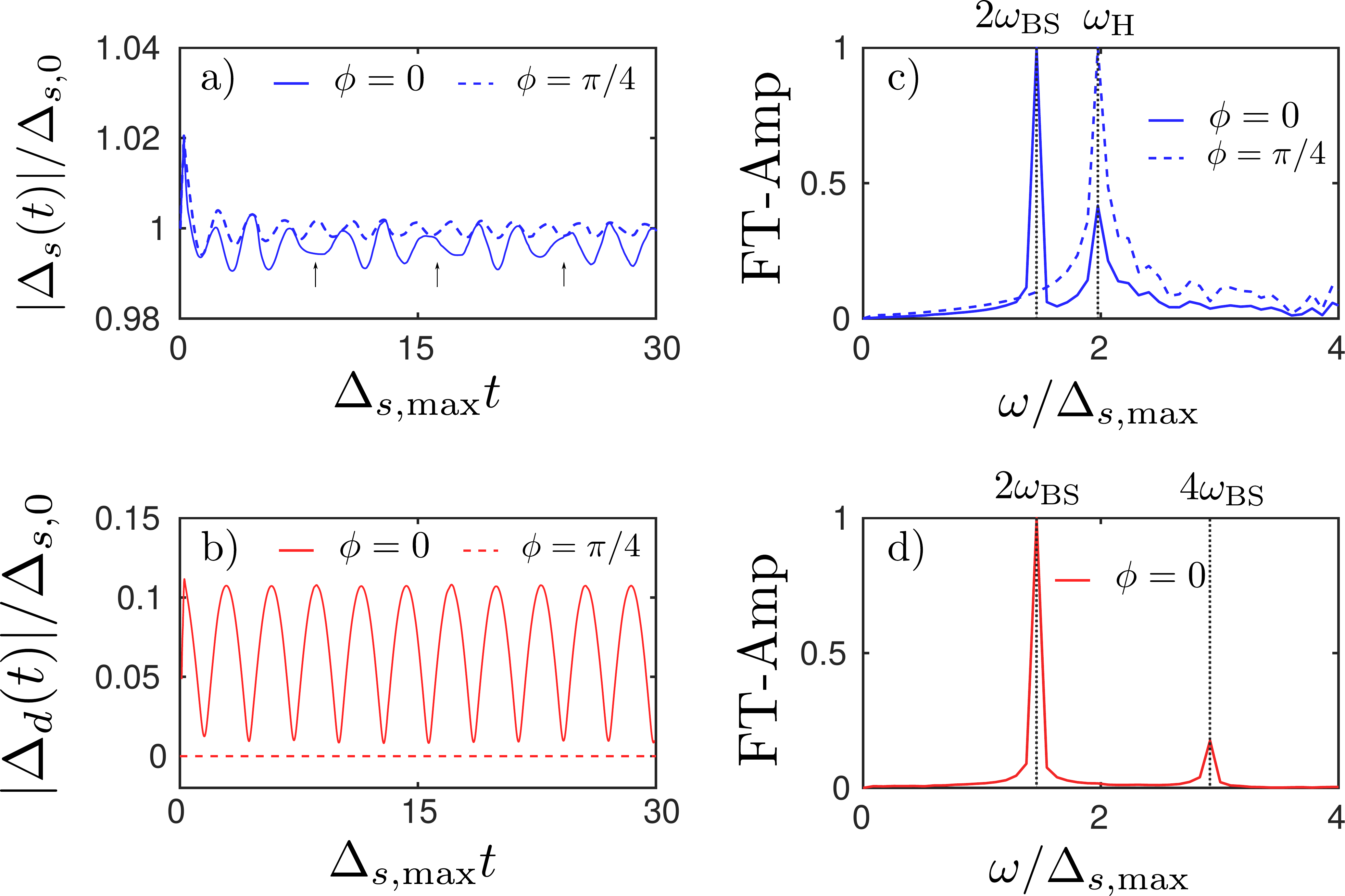}
	\caption{Calculated evolution of the superconducting gap amplitudes for an s-wave ground state and an applied light pulse with polarization at an angle $\phi = 0$ (solid line) and an angle $\phi = \pi/4$ (dashed line) to the $k_x$-axis. (a) and (b) show the short-time dynamics of the $s$- and $d_{x^2-y^2}$-wave order parameter, respectively. To make the existence of a second frequency in (a) clear, arrows illustrate the beating pattern. (c) and (d) refer to the Fourier transform of $|\Delta_s|$ and $|\Delta_{d_{x^2-y^2}}(t)|$, respectively. As $|\Delta_{d_{x^2-y^2}}(t)|$ remains $0$ for $\phi = \pi/4$, no Fourier transform of this case is shown in d). \label{fig:d_in_s}}
\end{figure}
In our case, the pseudospin expectations values $\langle \mathbf{s}_{\kk}\rangle$ at $t=0$ are the equilibrium values, with $s$- or $d_{x^2-y^2}$-wave symmetry. Consequently, one requires $B_{\kk,d_{x^2-y^2}}^z = \xi_{\kk,\mathbf{A},d_{x^2-y^2}}$ to be finite at non-zero t to induce a finite $d_{x^2-y^2}$-wave order parameter in the $s$-wave ground state and vice versa. In particular, a finite $\xi_{\kk,d_{x^2-y^2}} \sim \cos(A_x) - \cos(A_y)$ (see Supplementary Material for details) can be induced by a vector potential $\mathbf{A}$ with polarization at $\phi \approx 0$, where $\phi$ is the polar angle in momentum space. Importantly, the effect manifestly depends on the polarization of $\mathbf{A}$, suggesting a possibility to control the induced order parameter symmetry in non-equilibrium. As $\xi_{\kk,\Aq,d_{xy}},\xi_{\kk,\Aq,g_{xy(x^2-y^2)}}$ and the odd-parity components are identically zero, only $s$- and $d_{x^2-y^2}$-components remain in Eq. \eqref{eq:eom} leading to
\begin{align}\label{eq:eom_symc}
    \frac{d}{dt}\langle \mathbf{s}_{\kk,s} \rangle &= \mathbf{B}_{\kk,s} \times \langle \mathbf{s}_{\kk,s} \rangle + \mathbf{B}_{\kk,d_{x^2-y^2}}\times\langle \mathbf{s}_{\kk,d_{x^2-y^2}} \rangle,
    \nonumber\\
    \frac{d}{dt}\langle \mathbf{s}_{\kk,d_{x^2-y^2}} \rangle &= \mathbf{B}_{\kk,d_{x^2-y^2}} \times \langle \mathbf{s}_{\kk,s} \rangle + \mathbf{B}_{\kk,s}\times\langle \mathbf{s}_{\kk,d_{x^2-y^2}} \rangle.
 \end{align}
We note that this result does depend on the absence of spatial variations of ${\bf A}$. However, for THz light used in the experiments,\cite{Shimano2013} the wavelength is of the order $\sim$100 $\mu$m which is much larger than, e.g., superconducting coherence length that rarely exceeds $\sim$100 nm (as estimated by the upper critical field $H_{c2}$\cite{Yin2009,Shan2011,Griss2014}), justifying the assumption.

\begin{figure}	
	\includegraphics[width = 0.8\linewidth]{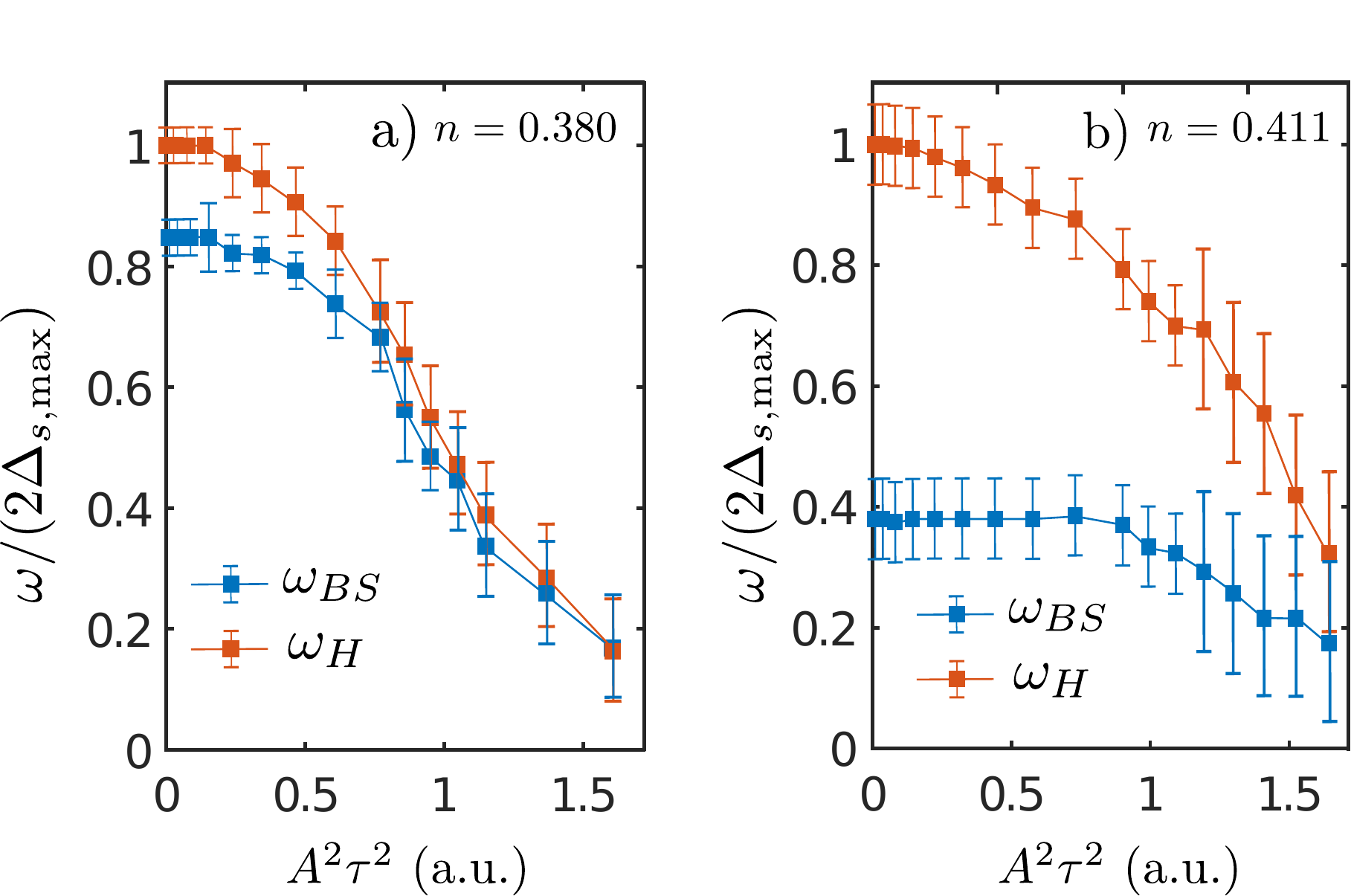}
	\caption{Fluence dependence of the Bardasis-Schrieffer and the Higgs mode in the $s$-wave ground state for  the filling (a) $n = 0.38$ and (b) $n = 0.411$, away (a) and close (b) to the phase transition to the $s+id_{x^2-y^2}$ ground state, respectively.  \label{fig:fluence}}
\end{figure}

We discuss now the short-time dynamics described by the equations \eqref{eq:eom_symc}. We integrate the equations numerically using Runge-Kutta method with a momentum grid of $513\times513$ points for fixed $V_s = V_d = -0.4t$ varying $n$ and the polarization of ${\bf A}(t)$. Let us consider first the situation of an extended $s$-wave ground state symmetry for $\Delta_\kk$, i.e. $\Delta_s \neq 0$ and $\Delta_{d_{x^2-y^2}} = 0$ at $n=0.411$. The transition into a $s + id_{x^2-y^2}$-state occurs at the filling $n \approx 0.415$.  We start the analysis by driving the model out of equilibrium with a vector potential in $k_x$-direction. The vector potential $\Aq$ is for simplicity chosen as $\Aq = \mathbf{A}_0\theta(t-\tau)\theta(2\tau-t)$ simulating two pulses at $t = \tau$ and $t = 2\tau$. Here we choose $\frac{e}{c}|\mathbf{A}_0| = 0.01$ and $\tau\Delta = 0.05$. Typically a Gaussian pulse shape is used in similar approaches\cite{Papenkort2007}, with a half-duration $t_{1/2} \ll \Delta^{-1}$ to work in the non-adiabatic regime. Our choice of the pulse is equivalent to this approach in the non-adiabatic limit, but we found it to be advantageous for computations. The results of the numerical calculations are shown in Fig. \ref{fig:d_in_s}. As one can see, the vector potential (pump pulse) induces a non-zero $d_{x^2-y^2}$-wave order parameter $\Delta_{d_{x^2-y^2}}$ in the extended $s$-wave ground state, which shows undamped oscillations at a frequency $\omega_{\text{BS}}$. Note that $|\Delta_{d_{x^2-y^2}}|$ shown in Fig. \ref{fig:d_in_s} oscillates at twice the frequency of $\Delta_{d_{x^2-y^2}}$ due to the absence of sign changes in the former. According to above considerations a finite $\Delta_{d_{x^2-y^2}}$ can only be induced if $\xi_{\kk,\mathbf{A},d_{x^2-y^2}}$ is finite for some period of time. Thus by applying the pulse along the diagonal $k_x = k_y$, for which $\xi_{\kk,d_{x^2-y^2}}$ remains zero, there is no induced oscillations of the $d_{x^2-y^2}$-wave order parameter. In both cases the $s$-wave order parameter oscillates with its Higgs mode $\omega_H = 2\Delta_{s,\text{max}}$, where $\Delta_{s,\text{max}} = \max\{\Delta_s\gamma_\kk | \kk \in FS\}$ is the maximum gap size on the Fermi surface. The frequency $\omega_{\text{BS}}$ is smaller than $\omega_H$ due to residual attractive interaction and therefore undamped, while $\omega_H$ shows the weak damping\cite{Barankov2004} observed for isotropic s-wave states. We attribute this to the gap being nearly isotropic along the Fermi surface at the doping we consider.
In Fig. \ref{fig:fluence} we show the fluence dependence of $\omega_{BS}$ and $\omega_H$. It can be seen that in the case where $\omega_{BS}$ and $\omega_H$ are well separated (Fig. \ref{fig:fluence} b) ), $\omega_{BS}$ is not directly affected by the light pulse, unlike $\omega_H$. However, once the Higgs mode frequency becomes close to $\omega_{BS}$, both start being suppressed by the increasing fluence and eventually become indistinguishable. Note that instead of using a pulse one could also by hand set a finite $\Delta_{d_{x^2-y^2}}\neq 0$ as initial value, as it was shown above that the pulse also generates a finite $d_{x^2-y^2}$-wave component. We find that the quench scenario yields results equivalent to Fig. \ref{fig:d_in_s}
%\subsection{Linear analysis of the equations}

\begin{figure}	
	\includegraphics[width = 0.8\linewidth]{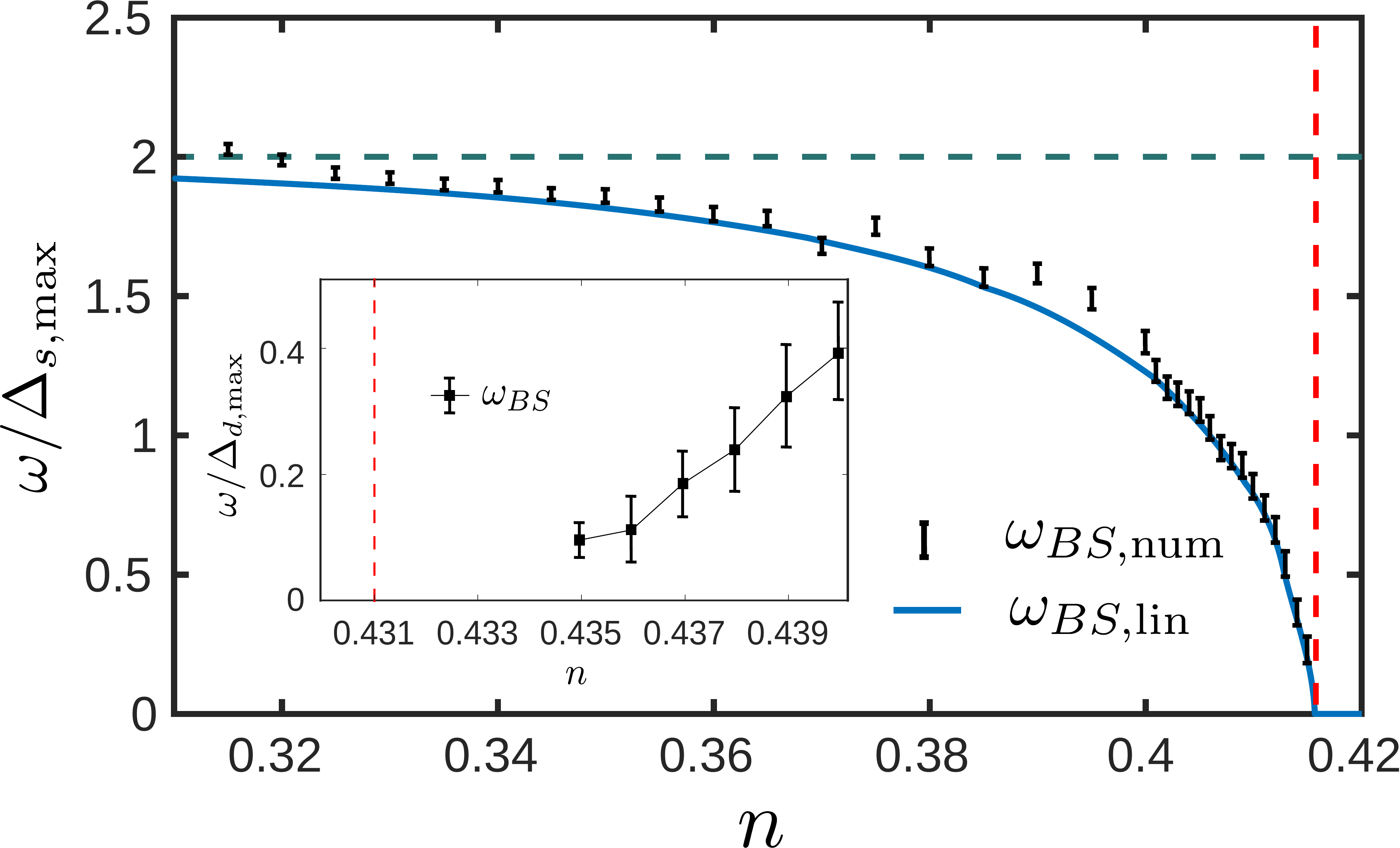}
	\caption{Frequency of the Bardasis-Schrieffer mode for different band fillings close to the transition point between extended $s$-wave and $d_{x^2-y^2}$-wave ground states at $n\approx0.415$. The numerical results (squares) and the results of the linearized equations (solid blue curve) are in good agreement. \label{fig:mode_dia}}
\end{figure}

To verify whether the oscillations at $\omega_{BS}$ are due to the existence of the Bardasis-Schrieffer mode in the sub-dominant pairing channel we analyze Eq. (\ref{eq:eom}) in the linear regime (i.e. weak driving) and compare the frequencies of the resulting modes to the ones observed in the non-equilibrium (see Fig. \ref{fig:d_in_s}). In particular, we linearize the Eq. \eqref{eq:eom} around the equilibrium state at $T=0$ and perform a subsequent Fourier transform ${-i\omega\delta\mathbf{s}_\kk = \mathbf{B}^{\text{eq}}_\kk\times\delta\mathbf{s}_\kk +  \delta\mathbf{B}_\kk\times\mathbf{s}^{\text{eq}}_\kk}$. Numerical solution of the homogeneous part of the resulting equations (see Supplementary Material for details) gives three types of solutions: (i) $\omega = 0$ mode for the ground state pairing symmetry independent of doping; (ii) $\omega = 2\Delta_{l,\text{max}}$ modes for the both types of the ground states $l=s$ and $l=d_{x^2-y^2}$ (Higgs mode),  and (iii) Bardasis-Schrieffer mode, which frequency depends on the proximity to the phase boundary between extended $s$-wave and $d_{x^2-y^2}$-wave symmetries in the equilibrium. We show the frequency of the Bardasis-Schrieffer mode in the $s$-wave ground state in Fig. \ref{fig:mode_dia} together with the results of the numerical solution of the full non-linear equation, Eq. (\ref{eq:eom}) as a function of doping. For values of $n$ far away from the transition this mode merges into the Higgs mode, while close to the phase transition it softens as is expected, signalling the transition to the new ground state. The results of linearized calculation of $\omega_{\text{BS}}$ are in good agreement with the results for the full non-linear dynamics.

We also considered the excitation of the Bardasis-Schrieffer mode in the $d_{x^2-y^2}$-wave ground state, which should also exist once the $s$-wave ground state is close enough. As the $d_{x^2-y^2}$-wave gap is nodal, all excited modes are strongly damped. However, we still see a clear damped mode at $\omega < 2\Delta_{d,max} = \omega_H$. Due to the finite size effects and the limited accuracy of the numerical integration, the error bar is quite broad. Therefore, we only show the frequency of this mode close to the $s+id$ transition in the inset of Fig. (\ref{fig:mode_dia}). 
To obtain these frequencies we can again equivalently use both, a pulse or a quench, for example, if one sets $\Delta_s = 0.1\Delta_{d_{x^2-y^2}}$. Note, for a better numerical accuracy  all frequencies in the inset of Fig. (\ref{fig:mode_dia}) are obtained via quench.
%\subsection{Oscillation of particlehas been observed in Raman spectroscopy of the iron-based superconductors\cite{Boehm18}.

\begin{figure}	
	\includegraphics[width = 0.9\linewidth]{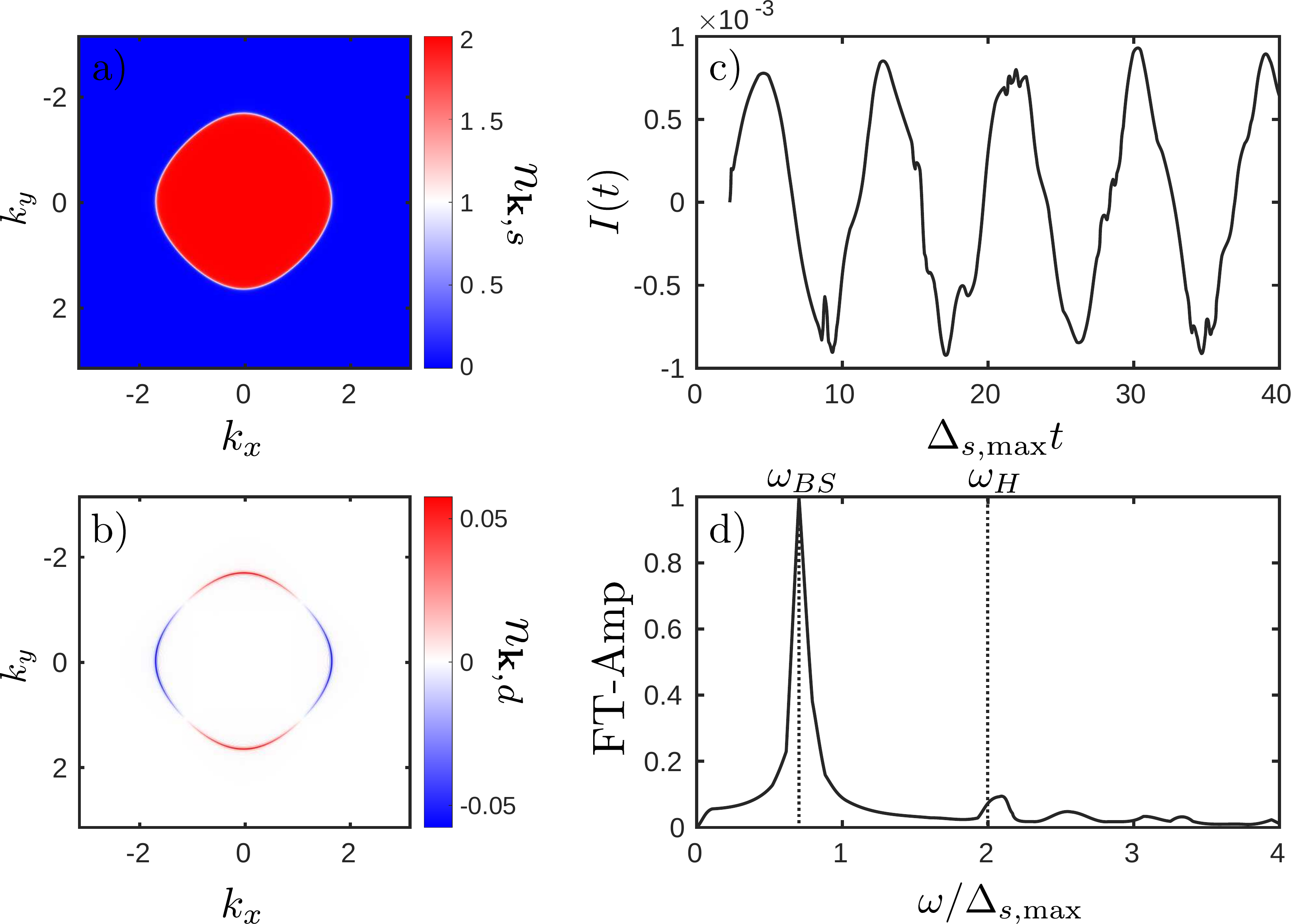}
	\caption{Oscillation of the quasiparticle distribution, projected into fully symmetric $s$-wave (a) and $d_{x^2-y^2}$-wave (b) parts. The oscillation of the $k_x$-integrated $d_{x^2-y^2}$-wave part in (c) shows a signature of the Bardasis-Schrieffer mode and of the Higgs mode as shown in its Fourier transform (d)).   \label{fig:fs_oscil}}
\end{figure}

The presence of the signatures of the Bardasis-Schrieffer mode in the order parameter dynamics raises the possibility of an enhanced third-harmonic generation when the pump frequency roughly matches $\omega_{\text{BS}}$. Unlike the Higgs mode, where the issues of quasiparticle contribution is still under debate,\cite{Cea2016,Cea2018,Tsuji2015,Matsunaga2017,Murotani2019,Kumar2019} the signature of a Bardasis-Schrieffer mode would be definitive as $\omega_{\text{BS}} < 2 \Delta_0$, below the edge of the quasiparticle continuum, at least for fully gapped $s$-wave superconductors.  At the same time, the amplitude oscillation of the order parameter (Higgs oscillation) was also predicted to show up in the tr-ARPES experiments.\cite{Nosarzewski2017} The quasiparticle distribution function that can be measured in these experiments can be addressed in our model.  In particular, the quasiparticle distribution function is determined by the $s_\kk^z$ component of the pseudospin $s_\kk^z = \frac{1}{2}\left(\langle c_{\kk\uparrow}^\dagger c_{\kk \uparrow}\rangle + \langle c_{\kk\downarrow}^\dagger c_{\kk \downarrow}\rangle -1 \right)$. In equilibrium  this quantity is equal to $-\xi_\kk/(2E_{\kk})$ and thus is $C_4$-symmetric. Due to the perturbation via an electric field, the tetragonal symmetry is temporarily broken down to $C_2$-symmetry and $s_\kk^z$ develops a finite $d_{x^2-y^2}$-wave component $s_\kk^{z,d}$, i.e. the quasiparticle distribution along $x$- and $y$-axis becomes asymmetric. Therefore, one expects that the non-equilibrium particle distribution acquires the information on $\omega_{\text{BS}}$ and $\omega_H$. To investigate this in detail we define the quantity $I(t) = \int_{-\pi}^{\pi}d k_x (2s_{\kk = (k_x,0)}^{z,d}(t))$,
which describes the integrated $d_{x^2-y^2}$-wave component of the quasiparticle distribution along the $k_y = 0$ cut and is equivalent to the number of particles with $k_y  = 0$. In Fig. \ref{fig:fs_oscil} we show $I(t)$ for the same parameters as in Fig. \ref{fig:d_in_s} for $\mathbf{A}$ oriented along $x$-axis. One can readily see that it oscillates mostly with $\omega = \omega_{\text{BS}}$, where we find the amplitude of the oscillations increases with an increasing pump strength. The integration along the $k_x = 0$ cut leads to similar results but shifted by a phase $\varphi = \pi$.

To conclude we analyze the short-time dynamics in single band unconventional superconductor with multiple competing pairing symmetries.  Driving the system out of equilibrium with a light pulse (modeled as a time-dependent vector potential), we show how the collective signatures of symmetries different from a given ground state symmetry, known as Bardasis-Schrieffer modes\cite{bardasis61,maiti15,maiti16,Boehm18} in the context of $s$-wave ground state in the equilibrium, evolve  as a function of doping and the polarization direction.   Depending on the polarization direction of the incoming light the tetragonal symmetry is broken, which necessarily induces a coupling of $s$- and $d$-wave symmetries in a non-equilibrium. Furthermore, we show that the particle distribution acquires an additional $d_{x^2-y^2}$-character, due to coupling to the vector potential and that this quantity shows a dominant signature of the Bardasis-Schrieffer mode frequency, which might be observed in time-resolved ARPES experiments. Therefore we conclude that taking the sub-dominant pairing channels and the corresponding interactions into account is important, when discussing polarization dependent excitation of unconventional superconductors.

{\it Acknowledgements}: M.A. M. and I.M. E. were supported by the
joint DFG-ANR Project (ER 463/8-1). P.A.V. acknowledges the support by the Rutgers University Center for Materials Theory Postdoctoral fellowship. I. P. is supported by the ANR grant (ANR-15-CE30-0025).

\bibliography{bibliography}

\begin{thebibliography}{43}%
\makeatletter
\providecommand \@ifxundefined [1]{%
 \@ifx{#1\undefined}
}%
\providecommand \@ifnum [1]{%
 \ifnum #1\expandafter \@firstoftwo
 \else \expandafter \@secondoftwo
 \fi
}%
\providecommand \@ifx [1]{%
 \ifx #1\expandafter \@firstoftwo
 \else \expandafter \@secondoftwo
 \fi
}%
\providecommand \natexlab [1]{#1}%
\providecommand \enquote  [1]{``#1''}%
\providecommand \bibnamefont  [1]{#1}%
\providecommand \bibfnamefont [1]{#1}%
\providecommand \citenamefont [1]{#1}%
\providecommand \href@noop [0]{\@secondoftwo}%
\providecommand \href [0]{\begingroup \@sanitize@url \@href}%
\providecommand \@href[1]{\@@startlink{#1}\@@href}%
\providecommand \@@href[1]{\endgroup#1\@@endlink}%
\providecommand \@sanitize@url [0]{\catcode `\\12\catcode `\$12\catcode
  `\&12\catcode `\#12\catcode `\^12\catcode `\_12\catcode `\%12\relax}%
\providecommand \@@startlink[1]{}%
\providecommand \@@endlink[0]{}%
\providecommand \url  [0]{\begingroup\@sanitize@url \@url }%
\providecommand \@url [1]{\endgroup\@href {#1}{\urlprefix }}%
\providecommand \urlprefix  [0]{URL }%
\providecommand \Eprint [0]{\href }%
\providecommand \doibase [0]{http://dx.doi.org/}%
\providecommand \selectlanguage [0]{\@gobble}%
\providecommand \bibinfo  [0]{\@secondoftwo}%
\providecommand \bibfield  [0]{\@secondoftwo}%
\providecommand \translation [1]{[#1]}%
\providecommand \BibitemOpen [0]{}%
\providecommand \bibitemStop [0]{}%
\providecommand \bibitemNoStop [0]{.\EOS\space}%
\providecommand \EOS [0]{\spacefactor3000\relax}%
\providecommand \BibitemShut  [1]{\csname bibitem#1\endcsname}%
\let\auto@bib@innerbib\@empty
%</preamble>
\bibitem [{\citenamefont {Pashkin}\ \emph {et~al.}(2010)\citenamefont
  {Pashkin}, \citenamefont {Porer}, \citenamefont {Beyer}, \citenamefont {Kim},
  \citenamefont {Dubroka}, \citenamefont {Bernhard}, \citenamefont {Yao},
  \citenamefont {Dagan}, \citenamefont {Hackl}, \citenamefont {Erb},
  \citenamefont {Demsar}, \citenamefont {Huber},\ and\ \citenamefont
  {Leitenstorfer}}]{Pashkin2010}%
  \BibitemOpen
  \bibfield  {author} {\bibinfo {author} {\bibfnamefont {A.}~\bibnamefont
  {Pashkin}}, \bibinfo {author} {\bibfnamefont {M.}~\bibnamefont {Porer}},
  \bibinfo {author} {\bibfnamefont {M.}~\bibnamefont {Beyer}}, \bibinfo
  {author} {\bibfnamefont {K.~W.}\ \bibnamefont {Kim}}, \bibinfo {author}
  {\bibfnamefont {A.}~\bibnamefont {Dubroka}}, \bibinfo {author} {\bibfnamefont
  {C.}~\bibnamefont {Bernhard}}, \bibinfo {author} {\bibfnamefont
  {X.}~\bibnamefont {Yao}}, \bibinfo {author} {\bibfnamefont {Y.}~\bibnamefont
  {Dagan}}, \bibinfo {author} {\bibfnamefont {R.}~\bibnamefont {Hackl}},
  \bibinfo {author} {\bibfnamefont {A.}~\bibnamefont {Erb}}, \bibinfo {author}
  {\bibfnamefont {J.}~\bibnamefont {Demsar}}, \bibinfo {author} {\bibfnamefont
  {R.}~\bibnamefont {Huber}}, \ and\ \bibinfo {author} {\bibfnamefont
  {A.}~\bibnamefont {Leitenstorfer}},\ }\href {\doibase
  https://doi.org/10.1103/PhysRevLett.105.067001} {\bibfield  {journal}
  {\bibinfo  {journal} {Phys. Rev. Lett.}\ }\textbf {\bibinfo {volume} {105}},\
  \bibinfo {pages} {067001} (\bibinfo {year} {2010})}\BibitemShut {NoStop}%
\bibitem [{\citenamefont {Beck}\ \emph {et~al.}(2011)\citenamefont {Beck},
  \citenamefont {Klammer}, \citenamefont {Lang}, \citenamefont {Leiderer},
  \citenamefont {Kabanov}, \citenamefont {Gol’tsman},\ and\ \citenamefont
  {Demsar}}]{Beck2011}%
  \BibitemOpen
  \bibfield  {author} {\bibinfo {author} {\bibfnamefont {M.}~\bibnamefont
  {Beck}}, \bibinfo {author} {\bibfnamefont {M.}~\bibnamefont {Klammer}},
  \bibinfo {author} {\bibfnamefont {S.}~\bibnamefont {Lang}}, \bibinfo {author}
  {\bibfnamefont {P.}~\bibnamefont {Leiderer}}, \bibinfo {author}
  {\bibfnamefont {V.~V.}\ \bibnamefont {Kabanov}}, \bibinfo {author}
  {\bibfnamefont {G.~N.}\ \bibnamefont {Gol’tsman}}, \ and\ \bibinfo {author}
  {\bibfnamefont {J.}~\bibnamefont {Demsar}},\ }\href {\doibase
  https://doi.org/10.1103/PhysRevLett.107.177007} {\bibfield  {journal}
  {\bibinfo  {journal} {Phys. Rev. Lett.}\ }\textbf {\bibinfo {volume} {107}},\
  \bibinfo {pages} {177007} (\bibinfo {year} {2011})}\BibitemShut {NoStop}%
\bibitem [{\citenamefont {Fausti}\ \emph {et~al.}(2011)\citenamefont {Fausti},
  \citenamefont {Tobey}, \citenamefont {Dean}, \citenamefont {Kaiser},
  \citenamefont {Dienst}, \citenamefont {Hoffmann}, \citenamefont {Pyon},
  \citenamefont {Takayama}, \citenamefont {Takagi},\ and\ \citenamefont
  {Cavalleri}}]{Cav2011Science}%
  \BibitemOpen
  \bibfield  {author} {\bibinfo {author} {\bibfnamefont {D.}~\bibnamefont
  {Fausti}}, \bibinfo {author} {\bibfnamefont {R.~I.}\ \bibnamefont {Tobey}},
  \bibinfo {author} {\bibfnamefont {N.}~\bibnamefont {Dean}}, \bibinfo {author}
  {\bibfnamefont {S.}~\bibnamefont {Kaiser}}, \bibinfo {author} {\bibfnamefont
  {A.}~\bibnamefont {Dienst}}, \bibinfo {author} {\bibfnamefont {M.~C.}\
  \bibnamefont {Hoffmann}}, \bibinfo {author} {\bibfnamefont {S.}~\bibnamefont
  {Pyon}}, \bibinfo {author} {\bibfnamefont {T.}~\bibnamefont {Takayama}},
  \bibinfo {author} {\bibfnamefont {H.}~\bibnamefont {Takagi}}, \ and\ \bibinfo
  {author} {\bibfnamefont {A.}~\bibnamefont {Cavalleri}},\ }\href {\doibase
  10.1126/science.1197294} {\bibfield  {journal} {\bibinfo  {journal}
  {Science}\ }\textbf {\bibinfo {volume} {331}},\ \bibinfo {pages} {189}
  (\bibinfo {year} {2011})}\BibitemShut {NoStop}%
\bibitem [{\citenamefont {Matsunaga}\ and\ \citenamefont
  {Shimano}(2012)}]{Matsunaga2012}%
  \BibitemOpen
  \bibfield  {author} {\bibinfo {author} {\bibfnamefont {R.}~\bibnamefont
  {Matsunaga}}\ and\ \bibinfo {author} {\bibfnamefont {R.}~\bibnamefont
  {Shimano}},\ }\href@noop {} {\bibfield  {journal} {\bibinfo  {journal} {Phys.
  Rev. Lett.}\ }\textbf {\bibinfo {volume} {109}},\ \bibinfo {pages} {187002}
  (\bibinfo {year} {2012})}\BibitemShut {NoStop}%
\bibitem [{\citenamefont {Dal~Conte}\ \emph {et~al.}(2012)\citenamefont
  {Dal~Conte}, \citenamefont {Giannetti}, \citenamefont {Coslovich},
  \citenamefont {Cilento}, \citenamefont {Bossini}, \citenamefont {Abebaw},
  \citenamefont {Banfi}, \citenamefont {Ferrini}, \citenamefont {Eisaki},
  \citenamefont {Greven}, \citenamefont {Damascelli}, \citenamefont {van~der
  Marel},\ and\ \citenamefont {Parmigiani}}]{Conte2012}%
  \BibitemOpen
  \bibfield  {author} {\bibinfo {author} {\bibfnamefont {S.}~\bibnamefont
  {Dal~Conte}}, \bibinfo {author} {\bibfnamefont {C.}~\bibnamefont
  {Giannetti}}, \bibinfo {author} {\bibfnamefont {G.}~\bibnamefont
  {Coslovich}}, \bibinfo {author} {\bibfnamefont {F.}~\bibnamefont {Cilento}},
  \bibinfo {author} {\bibfnamefont {D.}~\bibnamefont {Bossini}}, \bibinfo
  {author} {\bibfnamefont {T.}~\bibnamefont {Abebaw}}, \bibinfo {author}
  {\bibfnamefont {F.}~\bibnamefont {Banfi}}, \bibinfo {author} {\bibfnamefont
  {G.}~\bibnamefont {Ferrini}}, \bibinfo {author} {\bibfnamefont
  {H.}~\bibnamefont {Eisaki}}, \bibinfo {author} {\bibfnamefont
  {M.}~\bibnamefont {Greven}}, \bibinfo {author} {\bibfnamefont
  {A.}~\bibnamefont {Damascelli}}, \bibinfo {author} {\bibfnamefont
  {D.}~\bibnamefont {van~der Marel}}, \ and\ \bibinfo {author} {\bibfnamefont
  {F.}~\bibnamefont {Parmigiani}},\ }\href@noop {} {\bibfield  {journal}
  {\bibinfo  {journal} {Science}\ }\textbf {\bibinfo {volume} {335}},\ \bibinfo
  {pages} {1600} (\bibinfo {year} {2012})}\BibitemShut {NoStop}%
\bibitem [{\citenamefont {Beck}\ \emph {et~al.}(2013)\citenamefont {Beck},
  \citenamefont {Rousseau}, \citenamefont {Klammer}, \citenamefont {Leiderer},
  \citenamefont {Mittendorff}, \citenamefont {Winnerl}, \citenamefont {Helm},
  \citenamefont {Gol’tsman},\ and\ \citenamefont {Demsar}}]{Beck2013}%
  \BibitemOpen
  \bibfield  {author} {\bibinfo {author} {\bibfnamefont {M.}~\bibnamefont
  {Beck}}, \bibinfo {author} {\bibfnamefont {I.}~\bibnamefont {Rousseau}},
  \bibinfo {author} {\bibfnamefont {M.}~\bibnamefont {Klammer}}, \bibinfo
  {author} {\bibfnamefont {P.}~\bibnamefont {Leiderer}}, \bibinfo {author}
  {\bibfnamefont {M.}~\bibnamefont {Mittendorff}}, \bibinfo {author}
  {\bibfnamefont {S.}~\bibnamefont {Winnerl}}, \bibinfo {author} {\bibfnamefont
  {M.}~\bibnamefont {Helm}}, \bibinfo {author} {\bibfnamefont {G.~N.}\
  \bibnamefont {Gol’tsman}}, \ and\ \bibinfo {author} {\bibfnamefont
  {J.}~\bibnamefont {Demsar}},\ }\href {\doibase
  https://doi.org/10.1103/PhysRevLett.110.267003} {\bibfield  {journal}
  {\bibinfo  {journal} {Phys. Rev. Lett.}\ }\textbf {\bibinfo {volume} {110}},\
  \bibinfo {pages} {267003} (\bibinfo {year} {2013})}\BibitemShut {NoStop}%
\bibitem [{\citenamefont {Mansart}\ \emph {et~al.}(2013)\citenamefont
  {Mansart}, \citenamefont {Lorenzana}, \citenamefont {Mann}, \citenamefont
  {Odeh}, \citenamefont {Scarongella}, \citenamefont {Chergui},\ and\
  \citenamefont {Carbone}}]{Mansart2013}%
  \BibitemOpen
  \bibfield  {author} {\bibinfo {author} {\bibfnamefont {B.}~\bibnamefont
  {Mansart}}, \bibinfo {author} {\bibfnamefont {J.}~\bibnamefont {Lorenzana}},
  \bibinfo {author} {\bibfnamefont {A.}~\bibnamefont {Mann}}, \bibinfo {author}
  {\bibfnamefont {A.}~\bibnamefont {Odeh}}, \bibinfo {author} {\bibfnamefont
  {M.}~\bibnamefont {Scarongella}}, \bibinfo {author} {\bibfnamefont
  {M.}~\bibnamefont {Chergui}}, \ and\ \bibinfo {author} {\bibfnamefont
  {F.}~\bibnamefont {Carbone}},\ }\href {\doibase
  https://doi.org/10.1073/pnas.1218742110} {\bibfield  {journal} {\bibinfo
  {journal} {Proc. Natl. Acad. Sci. USA}\ }\textbf {\bibinfo {volume} {110}},\
  \bibinfo {pages} {4539} (\bibinfo {year} {2013})}\BibitemShut {NoStop}%
\bibitem [{\citenamefont {Matsunaga}\ \emph
  {et~al.}(2014{\natexlab{a}})\citenamefont {Matsunaga}, \citenamefont {Tsuji},
  \citenamefont {Fujita}, \citenamefont {Sugioka}, \citenamefont {Makise},
  \citenamefont {Uzawa}, \citenamefont {Terai}, \citenamefont {Wang},
  \citenamefont {Aoki},\ and\ \citenamefont {Shimano}}]{Shimano2014Science}%
  \BibitemOpen
  \bibfield  {author} {\bibinfo {author} {\bibfnamefont {R.}~\bibnamefont
  {Matsunaga}}, \bibinfo {author} {\bibfnamefont {N.}~\bibnamefont {Tsuji}},
  \bibinfo {author} {\bibfnamefont {H.}~\bibnamefont {Fujita}}, \bibinfo
  {author} {\bibfnamefont {A.}~\bibnamefont {Sugioka}}, \bibinfo {author}
  {\bibfnamefont {K.}~\bibnamefont {Makise}}, \bibinfo {author} {\bibfnamefont
  {Y.}~\bibnamefont {Uzawa}}, \bibinfo {author} {\bibfnamefont
  {H.}~\bibnamefont {Terai}}, \bibinfo {author} {\bibfnamefont
  {Z.}~\bibnamefont {Wang}}, \bibinfo {author} {\bibfnamefont {H.}~\bibnamefont
  {Aoki}}, \ and\ \bibinfo {author} {\bibfnamefont {R.}~\bibnamefont
  {Shimano}},\ }\href {\doibase 10.1126/science.1254697} {\bibfield  {journal}
  {\bibinfo  {journal} {Science}\ }\textbf {\bibinfo {volume} {345}},\ \bibinfo
  {pages} {1145} (\bibinfo {year} {2014}{\natexlab{a}})}\BibitemShut {NoStop}%
\bibitem [{\citenamefont {Hu}\ \emph {et~al.}(2014)\citenamefont {Hu},
  \citenamefont {Kaiser}, \citenamefont {Nicoletti}, \citenamefont {Hunt},
  \citenamefont {Gierz}, \citenamefont {Hoffmann}, \citenamefont {Le~Tacon},
  \citenamefont {Loew}, \citenamefont {Keimer},\ and\ \citenamefont
  {Cavalleri}}]{CavNature2014}%
  \BibitemOpen
  \bibfield  {author} {\bibinfo {author} {\bibfnamefont {W.}~\bibnamefont
  {Hu}}, \bibinfo {author} {\bibfnamefont {S.}~\bibnamefont {Kaiser}}, \bibinfo
  {author} {\bibfnamefont {D.}~\bibnamefont {Nicoletti}}, \bibinfo {author}
  {\bibfnamefont {C.~R.}\ \bibnamefont {Hunt}}, \bibinfo {author}
  {\bibfnamefont {I.}~\bibnamefont {Gierz}}, \bibinfo {author} {\bibfnamefont
  {M.~C.}\ \bibnamefont {Hoffmann}}, \bibinfo {author} {\bibfnamefont
  {M.}~\bibnamefont {Le~Tacon}}, \bibinfo {author} {\bibfnamefont
  {T.}~\bibnamefont {Loew}}, \bibinfo {author} {\bibfnamefont {B.}~\bibnamefont
  {Keimer}}, \ and\ \bibinfo {author} {\bibfnamefont {A.}~\bibnamefont
  {Cavalleri}},\ }\href {http://dx.doi.org/10.1038/nmat3963} {\bibfield
  {journal} {\bibinfo  {journal} {Nature Materials}\ }\textbf {\bibinfo
  {volume} {13}},\ \bibinfo {pages} {705 EP } (\bibinfo {year}
  {2014})}\BibitemShut {NoStop}%
\bibitem [{\citenamefont {Casandruc}\ \emph {et~al.}(2015)\citenamefont
  {Casandruc}, \citenamefont {Nicoletti}, \citenamefont {Rajasekaran},
  \citenamefont {Laplace}, \citenamefont {Khanna}, \citenamefont {Gu},
  \citenamefont {Hill},\ and\ \citenamefont {Cavalleri}}]{CavPRB2015}%
  \BibitemOpen
  \bibfield  {author} {\bibinfo {author} {\bibfnamefont {E.}~\bibnamefont
  {Casandruc}}, \bibinfo {author} {\bibfnamefont {D.}~\bibnamefont
  {Nicoletti}}, \bibinfo {author} {\bibfnamefont {S.}~\bibnamefont
  {Rajasekaran}}, \bibinfo {author} {\bibfnamefont {Y.}~\bibnamefont
  {Laplace}}, \bibinfo {author} {\bibfnamefont {V.}~\bibnamefont {Khanna}},
  \bibinfo {author} {\bibfnamefont {G.~D.}\ \bibnamefont {Gu}}, \bibinfo
  {author} {\bibfnamefont {J.~P.}\ \bibnamefont {Hill}}, \ and\ \bibinfo
  {author} {\bibfnamefont {A.}~\bibnamefont {Cavalleri}},\ }\href {\doibase
  10.1103/PhysRevB.91.174502} {\bibfield  {journal} {\bibinfo  {journal} {Phys.
  Rev. B}\ }\textbf {\bibinfo {volume} {91}},\ \bibinfo {pages} {174502}
  (\bibinfo {year} {2015})}\BibitemShut {NoStop}%
\bibitem [{\citenamefont {Matsunaga}\ \emph {et~al.}(2017)\citenamefont
  {Matsunaga}, \citenamefont {Tsuji}, \citenamefont {Makise}, \citenamefont
  {Terai}, \citenamefont {Aoki},\ and\ \citenamefont
  {Shimano}}]{Matsunaga2017}%
  \BibitemOpen
  \bibfield  {author} {\bibinfo {author} {\bibfnamefont {R.}~\bibnamefont
  {Matsunaga}}, \bibinfo {author} {\bibfnamefont {N.}~\bibnamefont {Tsuji}},
  \bibinfo {author} {\bibfnamefont {K.}~\bibnamefont {Makise}}, \bibinfo
  {author} {\bibfnamefont {H.}~\bibnamefont {Terai}}, \bibinfo {author}
  {\bibfnamefont {H.}~\bibnamefont {Aoki}}, \ and\ \bibinfo {author}
  {\bibfnamefont {R.}~\bibnamefont {Shimano}},\ }\href {\doibase
  https://doi.org/10.1103/PhysRevB.96.020505} {\bibfield  {journal} {\bibinfo
  {journal} {Phys. Rev. B}\ }\textbf {\bibinfo {volume} {96}},\ \bibinfo
  {pages} {020505(R)} (\bibinfo {year} {2017})}\BibitemShut {NoStop}%
\bibitem [{\citenamefont {Katsumi}\ \emph {et~al.}(2018)\citenamefont
  {Katsumi}, \citenamefont {Tsuji}, \citenamefont {Hamada}, \citenamefont
  {Matsunaga}, \citenamefont {Schneeloch}, \citenamefont {Zhong}, \citenamefont
  {Gu}, \citenamefont {Aoki}, \citenamefont {Gallais},\ and\ \citenamefont
  {Shimano}}]{Katsumi2018}%
  \BibitemOpen
  \bibfield  {author} {\bibinfo {author} {\bibfnamefont {K.}~\bibnamefont
  {Katsumi}}, \bibinfo {author} {\bibfnamefont {N.}~\bibnamefont {Tsuji}},
  \bibinfo {author} {\bibfnamefont {Y.~I.}\ \bibnamefont {Hamada}}, \bibinfo
  {author} {\bibfnamefont {R.}~\bibnamefont {Matsunaga}}, \bibinfo {author}
  {\bibfnamefont {J.}~\bibnamefont {Schneeloch}}, \bibinfo {author}
  {\bibfnamefont {R.~D.}\ \bibnamefont {Zhong}}, \bibinfo {author}
  {\bibfnamefont {G.~D.}\ \bibnamefont {Gu}}, \bibinfo {author} {\bibfnamefont
  {H.}~\bibnamefont {Aoki}}, \bibinfo {author} {\bibfnamefont {Y.}~\bibnamefont
  {Gallais}}, \ and\ \bibinfo {author} {\bibfnamefont {R.}~\bibnamefont
  {Shimano}},\ }\href {\doibase 10.1103/PhysRevLett.120.117001} {\bibfield
  {journal} {\bibinfo  {journal} {Phys. Rev. Lett.}\ }\textbf {\bibinfo
  {volume} {120}},\ \bibinfo {pages} {117001} (\bibinfo {year}
  {2018})}\BibitemShut {NoStop}%
\bibitem [{\citenamefont {Volkov}\ and\ \citenamefont
  {Kogan}(1974)}]{Volkov1974}%
  \BibitemOpen
  \bibfield  {author} {\bibinfo {author} {\bibfnamefont {A.~F.}\ \bibnamefont
  {Volkov}}\ and\ \bibinfo {author} {\bibfnamefont {S.~M.}\ \bibnamefont
  {Kogan}},\ }\href@noop {} {\bibfield  {journal} {\bibinfo  {journal} {Sov.
  Phys. JETP}\ }\textbf {\bibinfo {volume} {38}},\ \bibinfo {pages} {1018}
  (\bibinfo {year} {1974})}\BibitemShut {NoStop}%
\bibitem [{\citenamefont {Amin}\ \emph {et~al.}(2004)\citenamefont {Amin},
  \citenamefont {Bezuglyi}, \citenamefont {Kijko},\ and\ \citenamefont
  {Omelyanchouk}}]{Amin2004}%
  \BibitemOpen
  \bibfield  {author} {\bibinfo {author} {\bibfnamefont {M.}~\bibnamefont
  {Amin}}, \bibinfo {author} {\bibfnamefont {E.}~\bibnamefont {Bezuglyi}},
  \bibinfo {author} {\bibfnamefont {A.}~\bibnamefont {Kijko}}, \ and\ \bibinfo
  {author} {\bibfnamefont {A.}~\bibnamefont {Omelyanchouk}},\ }\href@noop {}
  {\bibfield  {journal} {\bibinfo  {journal} {Low Temp. Phys.}\ }\textbf
  {\bibinfo {volume} {30}},\ \bibinfo {pages} {661} (\bibinfo {year}
  {2004})}\BibitemShut {NoStop}%
\bibitem [{\citenamefont {Barankov}\ \emph {et~al.}(2004)\citenamefont
  {Barankov}, \citenamefont {Levitov},\ and\ \citenamefont
  {Spivak}}]{Barankov2004}%
  \BibitemOpen
  \bibfield  {author} {\bibinfo {author} {\bibfnamefont {R.~A.}\ \bibnamefont
  {Barankov}}, \bibinfo {author} {\bibfnamefont {L.~S.}\ \bibnamefont
  {Levitov}}, \ and\ \bibinfo {author} {\bibfnamefont {B.~Z.}\ \bibnamefont
  {Spivak}},\ }\href {\doibase 10.1103/PhysRevLett.93.160401} {\bibfield
  {journal} {\bibinfo  {journal} {Phys. Rev. Lett.}\ }\textbf {\bibinfo
  {volume} {93}},\ \bibinfo {pages} {160401} (\bibinfo {year}
  {2004})}\BibitemShut {NoStop}%
\bibitem [{\citenamefont {Yuzbashyan}\ \emph {et~al.}(2005)\citenamefont
  {Yuzbashyan}, \citenamefont {Altshuler}, \citenamefont {Kuznetsov},\ and\
  \citenamefont {Enolskii}}]{Yuzbashyan2005}%
  \BibitemOpen
  \bibfield  {author} {\bibinfo {author} {\bibfnamefont {E.}~\bibnamefont
  {Yuzbashyan}}, \bibinfo {author} {\bibfnamefont {B.}~\bibnamefont
  {Altshuler}}, \bibinfo {author} {\bibfnamefont {V.}~\bibnamefont
  {Kuznetsov}}, \ and\ \bibinfo {author} {\bibfnamefont {V.~Z.}\ \bibnamefont
  {Enolskii}},\ }\href@noop {} {\bibfield  {journal} {\bibinfo  {journal}
  {Phys. Rev. B}\ }\textbf {\bibinfo {volume} {72}},\ \bibinfo {pages} {220503}
  (\bibinfo {year} {2005})}\BibitemShut {NoStop}%
\bibitem [{\citenamefont {Yuzbashyan}\ \emph {et~al.}(2006)\citenamefont
  {Yuzbashyan}, \citenamefont {Tsyplyatyev},\ and\ \citenamefont
  {Altshuler}}]{Yuzbashyan2006}%
  \BibitemOpen
  \bibfield  {author} {\bibinfo {author} {\bibfnamefont {E.}~\bibnamefont
  {Yuzbashyan}}, \bibinfo {author} {\bibfnamefont {O.}~\bibnamefont
  {Tsyplyatyev}}, \ and\ \bibinfo {author} {\bibfnamefont {B.~L.}\ \bibnamefont
  {Altshuler}},\ }\href@noop {} {\bibfield  {journal} {\bibinfo  {journal}
  {Phys. Rev. Lett.}\ }\textbf {\bibinfo {volume} {96}},\ \bibinfo {pages}
  {097005} (\bibinfo {year} {2006})}\BibitemShut {NoStop}%
\bibitem [{\citenamefont {Barankov}\ and\ \citenamefont
  {Levitov}(2006)}]{Barankov2006}%
  \BibitemOpen
  \bibfield  {author} {\bibinfo {author} {\bibfnamefont {R.~A.}\ \bibnamefont
  {Barankov}}\ and\ \bibinfo {author} {\bibfnamefont {L.~S.}\ \bibnamefont
  {Levitov}},\ }\href {\doibase 10.1103/PhysRevLett.96.230403} {\bibfield
  {journal} {\bibinfo  {journal} {Phys. Rev. Lett.}\ }\textbf {\bibinfo
  {volume} {96}},\ \bibinfo {pages} {230403} (\bibinfo {year}
  {2006})}\BibitemShut {NoStop}%
\bibitem [{\citenamefont {Papenkort}\ \emph {et~al.}(2007)\citenamefont
  {Papenkort}, \citenamefont {Axt},\ and\ \citenamefont
  {Kuhn}}]{Papenkort2007}%
  \BibitemOpen
  \bibfield  {author} {\bibinfo {author} {\bibfnamefont {T.}~\bibnamefont
  {Papenkort}}, \bibinfo {author} {\bibfnamefont {V.}~\bibnamefont {Axt}}, \
  and\ \bibinfo {author} {\bibfnamefont {T.}~\bibnamefont {Kuhn}},\ }\href@noop
  {} {\bibfield  {journal} {\bibinfo  {journal} {Phys. Rev. B}\ }\textbf
  {\bibinfo {volume} {76}},\ \bibinfo {pages} {224522} (\bibinfo {year}
  {2007})}\BibitemShut {NoStop}%
\bibitem [{\citenamefont {Krull}\ \emph {et~al.}(2014)\citenamefont {Krull},
  \citenamefont {Manske}, \citenamefont {Uhrig},\ and\ \citenamefont
  {Schnyder}}]{Krull2014}%
  \BibitemOpen
  \bibfield  {author} {\bibinfo {author} {\bibfnamefont {H.}~\bibnamefont
  {Krull}}, \bibinfo {author} {\bibfnamefont {D.}~\bibnamefont {Manske}},
  \bibinfo {author} {\bibfnamefont {G.~S.}\ \bibnamefont {Uhrig}}, \ and\
  \bibinfo {author} {\bibfnamefont {A.~P.}\ \bibnamefont {Schnyder}},\ }\href
  {\doibase 10.1103/PhysRevB.90.014515} {\bibfield  {journal} {\bibinfo
  {journal} {Phys. Rev. B}\ }\textbf {\bibinfo {volume} {90}},\ \bibinfo
  {pages} {014515} (\bibinfo {year} {2014})}\BibitemShut {NoStop}%
\bibitem [{\citenamefont {Yuzbashyan}\ \emph {et~al.}(2015)\citenamefont
  {Yuzbashyan}, \citenamefont {Dzero}, \citenamefont {Gurarie},\ and\
  \citenamefont {Foster}}]{Big-Quench-Review2015}%
  \BibitemOpen
  \bibfield  {author} {\bibinfo {author} {\bibfnamefont {E.~A.}\ \bibnamefont
  {Yuzbashyan}}, \bibinfo {author} {\bibfnamefont {M.}~\bibnamefont {Dzero}},
  \bibinfo {author} {\bibfnamefont {V.}~\bibnamefont {Gurarie}}, \ and\
  \bibinfo {author} {\bibfnamefont {M.~S.}\ \bibnamefont {Foster}},\ }\href
  {\doibase 10.1103/PhysRevA.91.033628} {\bibfield  {journal} {\bibinfo
  {journal} {Phys. Rev. A}\ }\textbf {\bibinfo {volume} {91}},\ \bibinfo
  {pages} {033628} (\bibinfo {year} {2015})}\BibitemShut {NoStop}%
\bibitem [{\citenamefont {Tsuji}\ and\ \citenamefont {Aoki}(2015)}]{Tsuji2015}%
  \BibitemOpen
  \bibfield  {author} {\bibinfo {author} {\bibfnamefont {N.}~\bibnamefont
  {Tsuji}}\ and\ \bibinfo {author} {\bibfnamefont {H.}~\bibnamefont {Aoki}},\
  }\href@noop {} {\bibfield  {journal} {\bibinfo  {journal} {Phys. Rev. B}\
  }\textbf {\bibinfo {volume} {92}},\ \bibinfo {pages} {064508} (\bibinfo
  {year} {2015})}\BibitemShut {NoStop}%
\bibitem [{\citenamefont {Murotani}\ \emph {et~al.}(2017)\citenamefont
  {Murotani}, \citenamefont {Tsuji},\ and\ \citenamefont {Aoki}}]{Aoki-Higgs2}%
  \BibitemOpen
  \bibfield  {author} {\bibinfo {author} {\bibfnamefont {Y.}~\bibnamefont
  {Murotani}}, \bibinfo {author} {\bibfnamefont {N.}~\bibnamefont {Tsuji}}, \
  and\ \bibinfo {author} {\bibfnamefont {H.}~\bibnamefont {Aoki}},\ }\href
  {\doibase 10.1103/PhysRevB.95.104503} {\bibfield  {journal} {\bibinfo
  {journal} {Phys. Rev. B}\ }\textbf {\bibinfo {volume} {95}},\ \bibinfo
  {pages} {104503} (\bibinfo {year} {2017})}\BibitemShut {NoStop}%
\bibitem [{\citenamefont {Chou}\ \emph {et~al.}(2017)\citenamefont {Chou},
  \citenamefont {Liao},\ and\ \citenamefont {Foster}}]{Matt-Higgs}%
  \BibitemOpen
  \bibfield  {author} {\bibinfo {author} {\bibfnamefont {Y.-Z.}\ \bibnamefont
  {Chou}}, \bibinfo {author} {\bibfnamefont {Y.}~\bibnamefont {Liao}}, \ and\
  \bibinfo {author} {\bibfnamefont {M.~S.}\ \bibnamefont {Foster}},\ }\href
  {\doibase 10.1103/PhysRevB.95.104507} {\bibfield  {journal} {\bibinfo
  {journal} {Phys. Rev. B}\ }\textbf {\bibinfo {volume} {95}},\ \bibinfo
  {pages} {104507} (\bibinfo {year} {2017})}\BibitemShut {NoStop}%
\bibitem [{\citenamefont {Matsunaga}\ \emph
  {et~al.}(2014{\natexlab{b}})\citenamefont {Matsunaga}, \citenamefont {Tsuji},
  \citenamefont {Fujita}, \citenamefont {Sugioka}, \citenamefont {Makise},
  \citenamefont {Uzawa}, \citenamefont {Terai}, \citenamefont {Wang},
  \citenamefont {Aoki},\ and\ \citenamefont {Shimano}}]{Matsunaga2014}%
  \BibitemOpen
  \bibfield  {author} {\bibinfo {author} {\bibfnamefont {R.}~\bibnamefont
  {Matsunaga}}, \bibinfo {author} {\bibfnamefont {N.}~\bibnamefont {Tsuji}},
  \bibinfo {author} {\bibfnamefont {H.}~\bibnamefont {Fujita}}, \bibinfo
  {author} {\bibfnamefont {A.}~\bibnamefont {Sugioka}}, \bibinfo {author}
  {\bibfnamefont {K.}~\bibnamefont {Makise}}, \bibinfo {author} {\bibfnamefont
  {Y.}~\bibnamefont {Uzawa}}, \bibinfo {author} {\bibfnamefont
  {H.}~\bibnamefont {Terai}}, \bibinfo {author} {\bibfnamefont
  {Z.}~\bibnamefont {Wang}}, \bibinfo {author} {\bibfnamefont {H.}~\bibnamefont
  {Aoki}}, \ and\ \bibinfo {author} {\bibfnamefont {R.}~\bibnamefont
  {Shimano}},\ }\href@noop {} {\bibfield  {journal} {\bibinfo  {journal}
  {Science}\ }\textbf {\bibinfo {volume} {345}},\ \bibinfo {pages} {1145}
  (\bibinfo {year} {2014}{\natexlab{b}})}\BibitemShut {NoStop}%
\bibitem [{\citenamefont {Chu}\ \emph {et~al.}(2019)\citenamefont {Chu},
  \citenamefont {Kim}, \citenamefont {Katsumi}, \citenamefont {Kovalev},
  \citenamefont {Dawson}, \citenamefont {Schwarz}, \citenamefont {Yoshikawa},
  \citenamefont {Kim}, \citenamefont {Putzky}, \citenamefont {Li},
  \citenamefont {Raffy}, \citenamefont {Germanskiy}, \citenamefont {Deinert},
  \citenamefont {Awari}, \citenamefont {Ilyakov}, \citenamefont {Green},
  \citenamefont {Chen}, \citenamefont {Bawatna}, \citenamefont {Christiani},
  \citenamefont {Logvenov}, \citenamefont {Gallais}, \citenamefont {Boris},
  \citenamefont {Keimer}, \citenamefont {Schnyder}, \citenamefont {Manske},
  \citenamefont {Gensch}, \citenamefont {Wang}, \citenamefont {Shimano},\ and\
  \citenamefont {Kaiser}}]{Chu2019}%
  \BibitemOpen
  \bibfield  {author} {\bibinfo {author} {\bibfnamefont {H.}~\bibnamefont
  {Chu}}, \bibinfo {author} {\bibfnamefont {M.-J.}\ \bibnamefont {Kim}},
  \bibinfo {author} {\bibfnamefont {K.}~\bibnamefont {Katsumi}}, \bibinfo
  {author} {\bibfnamefont {S.}~\bibnamefont {Kovalev}}, \bibinfo {author}
  {\bibfnamefont {R.}~\bibnamefont {Dawson}}, \bibinfo {author} {\bibfnamefont
  {L.}~\bibnamefont {Schwarz}}, \bibinfo {author} {\bibfnamefont
  {N.}~\bibnamefont {Yoshikawa}}, \bibinfo {author} {\bibfnamefont
  {G.}~\bibnamefont {Kim}}, \bibinfo {author} {\bibfnamefont {D.}~\bibnamefont
  {Putzky}}, \bibinfo {author} {\bibfnamefont {Z.}~\bibnamefont {Li}}, \bibinfo
  {author} {\bibfnamefont {H.}~\bibnamefont {Raffy}}, \bibinfo {author}
  {\bibfnamefont {S.}~\bibnamefont {Germanskiy}}, \bibinfo {author}
  {\bibfnamefont {J.-C.}\ \bibnamefont {Deinert}}, \bibinfo {author}
  {\bibfnamefont {N.}~\bibnamefont {Awari}}, \bibinfo {author} {\bibfnamefont
  {I.}~\bibnamefont {Ilyakov}}, \bibinfo {author} {\bibfnamefont
  {B.}~\bibnamefont {Green}}, \bibinfo {author} {\bibfnamefont
  {M.}~\bibnamefont {Chen}}, \bibinfo {author} {\bibfnamefont {M.}~\bibnamefont
  {Bawatna}}, \bibinfo {author} {\bibfnamefont {G.}~\bibnamefont {Christiani}},
  \bibinfo {author} {\bibfnamefont {G.}~\bibnamefont {Logvenov}}, \bibinfo
  {author} {\bibfnamefont {Y.}~\bibnamefont {Gallais}}, \bibinfo {author}
  {\bibfnamefont {A.}~\bibnamefont {Boris}}, \bibinfo {author} {\bibfnamefont
  {B.}~\bibnamefont {Keimer}}, \bibinfo {author} {\bibfnamefont
  {A.}~\bibnamefont {Schnyder}}, \bibinfo {author} {\bibfnamefont
  {D.}~\bibnamefont {Manske}}, \bibinfo {author} {\bibfnamefont
  {M.}~\bibnamefont {Gensch}}, \bibinfo {author} {\bibfnamefont
  {Z.}~\bibnamefont {Wang}}, \bibinfo {author} {\bibfnamefont {R.}~\bibnamefont
  {Shimano}}, \ and\ \bibinfo {author} {\bibfnamefont {S.}~\bibnamefont
  {Kaiser}},\ }\href@noop {} {\bibfield  {journal} {\bibinfo  {journal}
  {arXiv:1901.06675}\ } (\bibinfo {year} {2019})}\BibitemShut {NoStop}%
\bibitem [{\citenamefont {Bardasis}\ and\ \citenamefont
  {Schrieffer}(1961)}]{bardasis61}%
  \BibitemOpen
  \bibfield  {author} {\bibinfo {author} {\bibfnamefont {A.}~\bibnamefont
  {Bardasis}}\ and\ \bibinfo {author} {\bibfnamefont {J.~R.}\ \bibnamefont
  {Schrieffer}},\ }\href {\doibase 10.1103/PhysRev.121.1050} {\bibfield
  {journal} {\bibinfo  {journal} {Phys. Rev.}\ }\textbf {\bibinfo {volume}
  {121}},\ \bibinfo {pages} {1050} (\bibinfo {year} {1961})}\BibitemShut
  {NoStop}%
\bibitem [{\citenamefont {Maiti}\ and\ \citenamefont
  {Hirschfeld}(2015)}]{maiti15}%
  \BibitemOpen
  \bibfield  {author} {\bibinfo {author} {\bibfnamefont {S.}~\bibnamefont
  {Maiti}}\ and\ \bibinfo {author} {\bibfnamefont {P.~J.}\ \bibnamefont
  {Hirschfeld}},\ }\href {\doibase 10.1103/PhysRevB.92.094506} {\bibfield
  {journal} {\bibinfo  {journal} {Phys. Rev. B}\ }\textbf {\bibinfo {volume}
  {92}},\ \bibinfo {pages} {094506} (\bibinfo {year} {2015})}\BibitemShut
  {NoStop}%
\bibitem [{\citenamefont {Maiti}\ \emph {et~al.}(2016)\citenamefont {Maiti},
  \citenamefont {Maier}, \citenamefont {B\"ohm}, \citenamefont {Hackl},\ and\
  \citenamefont {Hirschfeld}}]{maiti16}%
  \BibitemOpen
  \bibfield  {author} {\bibinfo {author} {\bibfnamefont {S.}~\bibnamefont
  {Maiti}}, \bibinfo {author} {\bibfnamefont {T.~A.}\ \bibnamefont {Maier}},
  \bibinfo {author} {\bibfnamefont {T.}~\bibnamefont {B\"ohm}}, \bibinfo
  {author} {\bibfnamefont {R.}~\bibnamefont {Hackl}}, \ and\ \bibinfo {author}
  {\bibfnamefont {P.~J.}\ \bibnamefont {Hirschfeld}},\ }\href {\doibase
  10.1103/PhysRevLett.117.257001} {\bibfield  {journal} {\bibinfo  {journal}
  {Phys. Rev. Lett.}\ }\textbf {\bibinfo {volume} {117}},\ \bibinfo {pages}
  {257001} (\bibinfo {year} {2016})}\BibitemShut {NoStop}%
\bibitem [{\citenamefont {B\"ohm}\ \emph {et~al.}(2018)\citenamefont {B\"ohm},
  \citenamefont {Kretzschmar}, \citenamefont {Baum}, \citenamefont {Rehm},
  \citenamefont {Jost}, \citenamefont {Ahangharnejhad}, \citenamefont
  {Thomale}, \citenamefont {Platt}, \citenamefont {Maier}, \citenamefont
  {Hanke}, \citenamefont {Moritz}, \citenamefont {Devereaux}, \citenamefont
  {Scalapino}, \citenamefont {Maiti}, \citenamefont {Hirschfeld}, \citenamefont
  {Adelmann}, \citenamefont {Wolf}, \citenamefont {Wen},\ and\ \citenamefont
  {Hackl}}]{Boehm18}%
  \BibitemOpen
  \bibfield  {author} {\bibinfo {author} {\bibfnamefont {T.}~\bibnamefont
  {B\"ohm}}, \bibinfo {author} {\bibfnamefont {F.}~\bibnamefont {Kretzschmar}},
  \bibinfo {author} {\bibfnamefont {A.}~\bibnamefont {Baum}}, \bibinfo {author}
  {\bibfnamefont {M.}~\bibnamefont {Rehm}}, \bibinfo {author} {\bibfnamefont
  {D.}~\bibnamefont {Jost}}, \bibinfo {author} {\bibfnamefont {R.~H.}\
  \bibnamefont {Ahangharnejhad}}, \bibinfo {author} {\bibfnamefont
  {R.}~\bibnamefont {Thomale}}, \bibinfo {author} {\bibfnamefont
  {C.}~\bibnamefont {Platt}}, \bibinfo {author} {\bibfnamefont {T.~A.}\
  \bibnamefont {Maier}}, \bibinfo {author} {\bibfnamefont {W.}~\bibnamefont
  {Hanke}}, \bibinfo {author} {\bibfnamefont {B.}~\bibnamefont {Moritz}},
  \bibinfo {author} {\bibfnamefont {T.~P.}\ \bibnamefont {Devereaux}}, \bibinfo
  {author} {\bibfnamefont {D.~J.}\ \bibnamefont {Scalapino}}, \bibinfo {author}
  {\bibfnamefont {S.}~\bibnamefont {Maiti}}, \bibinfo {author} {\bibfnamefont
  {P.~J.}\ \bibnamefont {Hirschfeld}}, \bibinfo {author} {\bibfnamefont
  {P.}~\bibnamefont {Adelmann}}, \bibinfo {author} {\bibfnamefont
  {T.}~\bibnamefont {Wolf}}, \bibinfo {author} {\bibfnamefont {H.-H.}\
  \bibnamefont {Wen}}, \ and\ \bibinfo {author} {\bibfnamefont
  {R.}~\bibnamefont {Hackl}},\ }\href {\doibase
  doi.org/10.1038/s41535-018-0118-z} {\bibfield  {journal} {\bibinfo  {journal}
  {npj Quantum Materials}\ }\textbf {\bibinfo {volume} {3}},\ \bibinfo {pages}
  {48} (\bibinfo {year} {2018})}\BibitemShut {NoStop}%
\bibitem [{\citenamefont {Davis}\ and\ \citenamefont {Lee}(2013)}]{Davis2013}%
  \BibitemOpen
  \bibfield  {author} {\bibinfo {author} {\bibfnamefont {J.~C.~S.}\
  \bibnamefont {Davis}}\ and\ \bibinfo {author} {\bibfnamefont {D.-H.}\
  \bibnamefont {Lee}},\ }\href {\doibase 10.1073/pnas.1316512110} {\bibfield
  {journal} {\bibinfo  {journal} {Proceedings of the National Academy of
  Sciences}\ }\textbf {\bibinfo {volume} {110}},\ \bibinfo {pages} {17623}
  (\bibinfo {year} {2013})}\BibitemShut {NoStop}%
\bibitem [{\citenamefont {M\"uller}\ \emph {et~al.}(2018)\citenamefont
  {M\"uller}, \citenamefont {Shen}, \citenamefont {Dzero},\ and\ \citenamefont
  {Eremin}}]{mueller18}%
  \BibitemOpen
  \bibfield  {author} {\bibinfo {author} {\bibfnamefont {M.~A.}\ \bibnamefont
  {M\"uller}}, \bibinfo {author} {\bibfnamefont {P.}~\bibnamefont {Shen}},
  \bibinfo {author} {\bibfnamefont {M.}~\bibnamefont {Dzero}}, \ and\ \bibinfo
  {author} {\bibfnamefont {I.}~\bibnamefont {Eremin}},\ }\href {\doibase
  10.1103/PhysRevB.98.024522} {\bibfield  {journal} {\bibinfo  {journal} {Phys.
  Rev. B}\ }\textbf {\bibinfo {volume} {98}},\ \bibinfo {pages} {024522}
  (\bibinfo {year} {2018})}\BibitemShut {NoStop}%
\bibitem [{\citenamefont {Kirmani}\ and\ \citenamefont
  {Dzero}(2019)}]{Kirmani2018}%
  \BibitemOpen
  \bibfield  {author} {\bibinfo {author} {\bibfnamefont {A.~A.}\ \bibnamefont
  {Kirmani}}\ and\ \bibinfo {author} {\bibfnamefont {M.}~\bibnamefont
  {Dzero}},\ }\href@noop {} {\bibfield  {journal} {\bibinfo  {journal} {Journal
  of Superconductivity and Novel Magnetism}\ } (\bibinfo {year}
  {2019})}\BibitemShut {NoStop}%
\bibitem [{\citenamefont {Anderson}(1958)}]{Anderson1958}%
  \BibitemOpen
  \bibfield  {author} {\bibinfo {author} {\bibfnamefont {P.~W.}\ \bibnamefont
  {Anderson}},\ }\href@noop {} {\bibfield  {journal} {\bibinfo  {journal}
  {Phys. Rev.}\ }\textbf {\bibinfo {volume} {112}},\ \bibinfo {pages} {1900}
  (\bibinfo {year} {1958})}\BibitemShut {NoStop}%
\bibitem [{\citenamefont {Matsunaga}\ \emph {et~al.}(2013)\citenamefont
  {Matsunaga}, \citenamefont {Hamada}, \citenamefont {Makise}, \citenamefont
  {Uzawa}, \citenamefont {Terai}, \citenamefont {Wang},\ and\ \citenamefont
  {Shimano}}]{Shimano2013}%
  \BibitemOpen
  \bibfield  {author} {\bibinfo {author} {\bibfnamefont {R.}~\bibnamefont
  {Matsunaga}}, \bibinfo {author} {\bibfnamefont {Y.~I.}\ \bibnamefont
  {Hamada}}, \bibinfo {author} {\bibfnamefont {K.}~\bibnamefont {Makise}},
  \bibinfo {author} {\bibfnamefont {Y.}~\bibnamefont {Uzawa}}, \bibinfo
  {author} {\bibfnamefont {H.}~\bibnamefont {Terai}}, \bibinfo {author}
  {\bibfnamefont {Z.}~\bibnamefont {Wang}}, \ and\ \bibinfo {author}
  {\bibfnamefont {R.}~\bibnamefont {Shimano}},\ }\href@noop {} {\bibfield
  {journal} {\bibinfo  {journal} {Phys. Rev. Lett.}\ }\textbf {\bibinfo
  {volume} {111}},\ \bibinfo {pages} {057002} (\bibinfo {year}
  {2013})}\BibitemShut {NoStop}%
\bibitem [{\citenamefont {Yin}\ \emph {et~al.}(2009)\citenamefont {Yin},
  \citenamefont {Zech}, \citenamefont {Williams}, \citenamefont {Wang},
  \citenamefont {Wu}, \citenamefont {Chen},\ and\ \citenamefont
  {Hoffman}}]{Yin2009}%
  \BibitemOpen
  \bibfield  {author} {\bibinfo {author} {\bibfnamefont {Y.}~\bibnamefont
  {Yin}}, \bibinfo {author} {\bibfnamefont {M.}~\bibnamefont {Zech}}, \bibinfo
  {author} {\bibfnamefont {T.~L.}\ \bibnamefont {Williams}}, \bibinfo {author}
  {\bibfnamefont {X.~F.}\ \bibnamefont {Wang}}, \bibinfo {author}
  {\bibfnamefont {G.}~\bibnamefont {Wu}}, \bibinfo {author} {\bibfnamefont
  {X.~H.}\ \bibnamefont {Chen}}, \ and\ \bibinfo {author} {\bibfnamefont
  {J.~E.}\ \bibnamefont {Hoffman}},\ }\href {\doibase
  10.1103/PhysRevLett.102.097002} {\bibfield  {journal} {\bibinfo  {journal}
  {Phys. Rev. Lett.}\ }\textbf {\bibinfo {volume} {102}},\ \bibinfo {pages}
  {097002} (\bibinfo {year} {2009})}\BibitemShut {NoStop}%
\bibitem [{\citenamefont {Shan}\ \emph {et~al.}(2011)\citenamefont {Shan},
  \citenamefont {Wang}, \citenamefont {Shen}, \citenamefont {Zeng},
  \citenamefont {Huang}, \citenamefont {Li}, \citenamefont {Wang},
  \citenamefont {Yang}, \citenamefont {Ren}, \citenamefont {Wang},
  \citenamefont {Pan},\ and\ \citenamefont {Wen}}]{Shan2011}%
  \BibitemOpen
  \bibfield  {author} {\bibinfo {author} {\bibfnamefont {L.}~\bibnamefont
  {Shan}}, \bibinfo {author} {\bibfnamefont {Y.-L.}\ \bibnamefont {Wang}},
  \bibinfo {author} {\bibfnamefont {B.}~\bibnamefont {Shen}}, \bibinfo {author}
  {\bibfnamefont {B.}~\bibnamefont {Zeng}}, \bibinfo {author} {\bibfnamefont
  {Y.}~\bibnamefont {Huang}}, \bibinfo {author} {\bibfnamefont
  {A.}~\bibnamefont {Li}}, \bibinfo {author} {\bibfnamefont {D.}~\bibnamefont
  {Wang}}, \bibinfo {author} {\bibfnamefont {H.}~\bibnamefont {Yang}}, \bibinfo
  {author} {\bibfnamefont {C.}~\bibnamefont {Ren}}, \bibinfo {author}
  {\bibfnamefont {Q.-H.}\ \bibnamefont {Wang}}, \bibinfo {author}
  {\bibfnamefont {S.~H.}\ \bibnamefont {Pan}}, \ and\ \bibinfo {author}
  {\bibfnamefont {H.-H.}\ \bibnamefont {Wen}},\ }\href
  {https://doi.org/10.1038/nphys1908} {\bibfield  {journal} {\bibinfo
  {journal} {Nature Physics}\ }\textbf {\bibinfo {volume} {7}},\ \bibinfo
  {pages} {325 EP } (\bibinfo {year} {2011})}\BibitemShut {NoStop}%
\bibitem [{\citenamefont {Grissonnanche}\ \emph {et~al.}(2014)\citenamefont
  {Grissonnanche}, \citenamefont {Cyr-Choinière}, \citenamefont {Laliberté},
  \citenamefont {René~de Cotret}, \citenamefont {Juneau-Fecteau},
  \citenamefont {Dufour-Beauséjour}, \citenamefont {Delage}, \citenamefont
  {LeBoeuf}, \citenamefont {Chang}, \citenamefont {Ramshaw}, \citenamefont
  {Bonn}, \citenamefont {Hardy}, \citenamefont {Liang}, \citenamefont {Adachi},
  \citenamefont {Hussey}, \citenamefont {Vignolle}, \citenamefont {Proust},
  \citenamefont {Sutherland}, \citenamefont {Krämer}, \citenamefont {Park},
  \citenamefont {Graf}, \citenamefont {Doiron-Leyraud},\ and\ \citenamefont
  {Taillefer}}]{Griss2014}%
  \BibitemOpen
  \bibfield  {author} {\bibinfo {author} {\bibfnamefont {G.}~\bibnamefont
  {Grissonnanche}}, \bibinfo {author} {\bibfnamefont {O.}~\bibnamefont
  {Cyr-Choinière}}, \bibinfo {author} {\bibfnamefont {F.}~\bibnamefont
  {Laliberté}}, \bibinfo {author} {\bibfnamefont {S.}~\bibnamefont {René~de
  Cotret}}, \bibinfo {author} {\bibfnamefont {A.}~\bibnamefont
  {Juneau-Fecteau}}, \bibinfo {author} {\bibfnamefont {S.}~\bibnamefont
  {Dufour-Beauséjour}}, \bibinfo {author} {\bibfnamefont {M.~.}\ \bibnamefont
  {Delage}}, \bibinfo {author} {\bibfnamefont {D.}~\bibnamefont {LeBoeuf}},
  \bibinfo {author} {\bibfnamefont {J.}~\bibnamefont {Chang}}, \bibinfo
  {author} {\bibfnamefont {B.~J.}\ \bibnamefont {Ramshaw}}, \bibinfo {author}
  {\bibfnamefont {D.~A.}\ \bibnamefont {Bonn}}, \bibinfo {author}
  {\bibfnamefont {W.~N.}\ \bibnamefont {Hardy}}, \bibinfo {author}
  {\bibfnamefont {R.}~\bibnamefont {Liang}}, \bibinfo {author} {\bibfnamefont
  {S.}~\bibnamefont {Adachi}}, \bibinfo {author} {\bibfnamefont {N.~E.}\
  \bibnamefont {Hussey}}, \bibinfo {author} {\bibfnamefont {B.}~\bibnamefont
  {Vignolle}}, \bibinfo {author} {\bibfnamefont {C.}~\bibnamefont {Proust}},
  \bibinfo {author} {\bibfnamefont {M.}~\bibnamefont {Sutherland}}, \bibinfo
  {author} {\bibfnamefont {S.}~\bibnamefont {Krämer}}, \bibinfo {author}
  {\bibfnamefont {J.~H.}\ \bibnamefont {Park}}, \bibinfo {author}
  {\bibfnamefont {D.}~\bibnamefont {Graf}}, \bibinfo {author} {\bibfnamefont
  {N.}~\bibnamefont {Doiron-Leyraud}}, \ and\ \bibinfo {author} {\bibfnamefont
  {L.}~\bibnamefont {Taillefer}},\ }\href {\doibase 10.1038/ncomms4280}
  {\bibfield  {journal} {\bibinfo  {journal} {Nature Commun.}\ }\textbf
  {\bibinfo {volume} {5}},\ \bibinfo {pages} {3280} (\bibinfo {year}
  {2014})}\BibitemShut {NoStop}%
\bibitem [{\citenamefont {Cea}\ \emph {et~al.}(2016)\citenamefont {Cea},
  \citenamefont {Castellani},\ and\ \citenamefont {Benfatto}}]{Cea2016}%
  \BibitemOpen
  \bibfield  {author} {\bibinfo {author} {\bibfnamefont {T.}~\bibnamefont
  {Cea}}, \bibinfo {author} {\bibfnamefont {C.}~\bibnamefont {Castellani}}, \
  and\ \bibinfo {author} {\bibfnamefont {L.}~\bibnamefont {Benfatto}},\ }\href
  {\doibase 10.1103/PhysRevB.93.180507} {\bibfield  {journal} {\bibinfo
  {journal} {Phys. Rev. B}\ }\textbf {\bibinfo {volume} {93}},\ \bibinfo
  {pages} {180507} (\bibinfo {year} {2016})}\BibitemShut {NoStop}%
\bibitem [{\citenamefont {Cea}\ \emph {et~al.}(2018)\citenamefont {Cea},
  \citenamefont {Barone}, \citenamefont {Castellani},\ and\ \citenamefont
  {Benfatto}}]{Cea2018}%
  \BibitemOpen
  \bibfield  {author} {\bibinfo {author} {\bibfnamefont {T.}~\bibnamefont
  {Cea}}, \bibinfo {author} {\bibfnamefont {P.}~\bibnamefont {Barone}},
  \bibinfo {author} {\bibfnamefont {C.}~\bibnamefont {Castellani}}, \ and\
  \bibinfo {author} {\bibfnamefont {L.}~\bibnamefont {Benfatto}},\ }\href
  {\doibase 10.1103/PhysRevB.97.094516} {\bibfield  {journal} {\bibinfo
  {journal} {Phys. Rev. B}\ }\textbf {\bibinfo {volume} {97}},\ \bibinfo
  {pages} {094516} (\bibinfo {year} {2018})}\BibitemShut {NoStop}%
\bibitem [{\citenamefont {Murotani}\ and\ \citenamefont
  {Shimano}(2019)}]{Murotani2019}%
  \BibitemOpen
  \bibfield  {author} {\bibinfo {author} {\bibfnamefont {Y.}~\bibnamefont
  {Murotani}}\ and\ \bibinfo {author} {\bibfnamefont {R.}~\bibnamefont
  {Shimano}},\ }\href {\doibase 10.1103/PhysRevB.99.224510} {\bibfield
  {journal} {\bibinfo  {journal} {Phys. Rev. B}\ }\textbf {\bibinfo {volume}
  {99}},\ \bibinfo {pages} {224510} (\bibinfo {year} {2019})}\BibitemShut
  {NoStop}%
\bibitem [{\citenamefont {Kumar}\ and\ \citenamefont
  {Kemper}(2019)}]{Kumar2019}%
  \BibitemOpen
  \bibfield  {author} {\bibinfo {author} {\bibfnamefont {A.}~\bibnamefont
  {Kumar}}\ and\ \bibinfo {author} {\bibfnamefont {A.}~\bibnamefont {Kemper}},\
  }\href {https://arxiv.org/abs/1902.09549} {\bibfield  {journal} {\bibinfo
  {journal} {arXiv:1902.09549}\ } (\bibinfo {year} {2019})}\BibitemShut
  {NoStop}%
\bibitem [{\citenamefont {Nosarzewski}\ \emph {et~al.}(2017)\citenamefont
  {Nosarzewski}, \citenamefont {Moritz}, \citenamefont {Freericks},
  \citenamefont {Kemper},\ and\ \citenamefont {Devereaux}}]{Nosarzewski2017}%
  \BibitemOpen
  \bibfield  {author} {\bibinfo {author} {\bibfnamefont {B.}~\bibnamefont
  {Nosarzewski}}, \bibinfo {author} {\bibfnamefont {B.}~\bibnamefont {Moritz}},
  \bibinfo {author} {\bibfnamefont {J.~K.}\ \bibnamefont {Freericks}}, \bibinfo
  {author} {\bibfnamefont {A.~F.}\ \bibnamefont {Kemper}}, \ and\ \bibinfo
  {author} {\bibfnamefont {T.~P.}\ \bibnamefont {Devereaux}},\ }\href {\doibase
  10.1103/PhysRevB.96.184518} {\bibfield  {journal} {\bibinfo  {journal} {Phys.
  Rev. B}\ }\textbf {\bibinfo {volume} {96}},\ \bibinfo {pages} {184518}
  (\bibinfo {year} {2017})}\BibitemShut {NoStop}%
\end{thebibliography}%

\newpage

%\documentclass[aps,prb,twocolumn,floatfix]{revtex4}
%
%\pdfoutput=1
%
%
%\usepackage{times}
%\usepackage{amsfonts}
%\usepackage{amssymb}
%\usepackage{amsmath}
%\usepackage{graphicx}
%\usepackage{bm}
%\usepackage{verbatim}
%
%\usepackage{bm}
%\usepackage{color}
%\usepackage{stackrel}
%\usepackage{accents}
%\usepackage{hyperref}
%\usepackage{latexsym}
%
%\usepackage{ulem}
%
%
%
%\newcommand{\kk}{\mathbf{k}}
%\newcommand{\Aq}{\mathbf{A}}
%\newcommand{\gs}{\gamma_{\kk,s}}
%\newcommand{\gd}{\gamma_{\kk,d_{x^2-y^2}}}
%\newcommand{\PV}[1]{\textcolor{red}{#1}}
%\newcommand{\MM}[1]{\textcolor{blue}{#1}}
%
%%%%%%%%%% Prefix a "S" to all equations, figures, tables and reset the counter %%%%%%%%%%
\setcounter{equation}{0}
\setcounter{figure}{0}
\setcounter{table}{0}
\makeatletter
\renewcommand{\theequation}{S\arabic{equation}}
\renewcommand{\thefigure}{S\arabic{figure}}
\renewcommand{\bibnumfmt}[1]{[S#1]}
\renewcommand{\citenumfont}[1]{S#1}
%%%%%%%%%%% Prefix a "S" to all equations, figures, tables and reset the counter %%%%%%%%%%
%
%
%%\makeatletter
%
%%\makeatother
%
%\begin{document}
	
	\title{Collective modes in pumped unconventional superconductors with competing ground states: Supplementary Material}
	
	\author{Marvin A. M\"{u}ller$^{1}$, Pavel A. Volkov$^{1,2}$, Indranil Paul$^3$,  and Ilya M. Eremin$^1$}
	\affiliation{$^1$ Institut f\"{u}r Theoretische Physik III, Ruhr-Universit\"{a}t Bochum, D-44801 Bochum, Germany \\
		$^2$Department of Physics and Astronomy, Center for Materials Theory, Rutgers University, Piscataway, NJ 08854 USA \\
		$^3$Laboratoire Mat\'{e}riaux et Ph\'{e}nom\`{e}nes Quantiques, Universit\'{e} de Paris, CNRS, F-75013, Paris, France}

\onecolumngrid
\setcounter{page}{1}
\begin{center}
\textbf{\large Collective modes in pumped unconventional superconductors with competing ground states: Supplementary Material} 
\end{center}
\begin{center}
Marvin A. M\"{u}ller$^{1}$, Pavel A. Volkov$^{1,2}$, Indranil Paul$^3$,  and Ilya M. Eremin$^1$\\
\textit{\small $^1$ Institut f\"{u}r Theoretische Physik III, Ruhr-Universit\"{a}t Bochum, D-44801 Bochum, Germany \\
$^2$Department of Physics and Astronomy, Center for Materials Theory, Rutgers University, Piscataway, NJ 08854 USA \\
$^3$Laboratoire Mat\'{e}riaux et Ph\'{e}nom\`{e}nes Quantiques, Universit\'{e} de Paris, CNRS, F-75013, Paris, France}
\end{center}

	\section{Symmetrized Equations of Motion}
	Here we give detailed information about the derivation of the equations of motion and its symmetrization. After performing a minimal substitution to include the vector potential the Hamiltonian has the following form
		\begin{align}
		H = \sum_\kk \mathbf{B}_\kk\cdot \mathbf{s}_\kk + (\xi_{\kk + \frac{e}{c}\mathbf{A}} - \xi_{\kk - \frac{e}{c}\mathbf{A}})(c_{\kk\uparrow}^\dagger c_{\kk\uparrow} + c_{-\kk\downarrow} c_{-\kk\downarrow}^\dagger),
		\end{align}
		where ${\mathbf{B}_\kk = \left(2\Delta^\prime_\kk,2\Delta^{\prime\prime}_\kk,\xi_{\kk + \frac{e}{c}\mathbf{A}} + \xi_{\kk - \frac{e}{c}\mathbf{A}}\right)^T}$. The last term transforms as $\frac{1}{2} \left( c_{\kk\uparrow}^\dagger,  c_{-\kk\downarrow}\right) \sigma_0 \begin{pmatrix} c_{\kk\uparrow} \\ c_{-\kk\downarrow}^\dagger\end{pmatrix}$, where $\sigma_0$ is the unit matrix and thus commutes with all other pseudospins. Therefore the structure of the equations of motions remains unaffected
		\begin{align}\label{eq:eom_supp}
		\frac{d}{dt}\langle\mathbf{s}_\kk\rangle = \frac{i}{\hbar} \langle [H,\mathbf{s}_\kk]\rangle = \mathbf{B}_\kk \times \langle\mathbf{s}_\kk\rangle.
		\end{align}\\
	Now one can rewrite $\mathbf{s}_\kk$ and $\mathbf{B}_\kk$ in Eq. \eqref{eq:eom_supp} as a sum of all even parity irreducible representations of the tetragonal $D_{4h}$ symmetry group as discussed in the main text. Using the product rules for the irreducible representations of the $D_{4h}$ symmetry group one can split this equations into four orthogonal parts
	\begin{align}\label{eq:eom_full_supp}
	\frac{d}{dt}\langle \mathbf{s}_{\kk,s} \rangle &= \mathbf{B}_{\kk,s} \times \langle \mathbf{s}_{\kk,s} \rangle + \mathbf{B}_{\kk,d_{x^2-y^2}}\times\langle \mathbf{s}_{\kk,d_{x^2-y^2}} \rangle +
	\mathbf{B}_{\kk,d_{xy}} \times \langle \mathbf{s}_{\kk,d_{xy}} \rangle + \mathbf{B}_{\kk,g_{xy(x^2-y^2)}}\times\langle \mathbf{s}_{\kk,g_{xy(x^2-y^2)}} \rangle,
	\nonumber\\
	\frac{d}{dt}\langle \mathbf{s}_{\kk,d_{x^2-y^2}} \rangle &= \mathbf{B}_{\kk,d_{x^2-y^2}} \times \langle \mathbf{s}_{\kk,s} \rangle + \mathbf{B}_{\kk,s}\times\langle \mathbf{s}_{\kk,d_{x^2-y^2}} \rangle + \mathbf{B}_{\kk,g_{xy(x^2-y^2)}} \times \langle \mathbf{s}_{\kk,d_{xy}} \rangle + \mathbf{B}_{\kk,d_{xy}}\times\langle \mathbf{s}_{\kk,g_{xy(x^2-y^2)}} \rangle, \nonumber\\
	\frac{d}{dt}\langle \mathbf{s}_{\kk,d_{xy}} \rangle &= \mathbf{B}_{\kk,s} \times \langle \mathbf{s}_{\kk,s} \rangle + \mathbf{B}_{\kk,g_{xy(x^2-y^2)}} \times  \langle \mathbf{s}_{\kk,d_{x^2-y^2}} \rangle + \mathbf{B}_{\kk,d_{x^2-y^2}}\times\langle \mathbf{s}_{\kk,g_{xy(x^2-y^2)}} \rangle, \nonumber\\
	\frac{d}{dt}\langle \mathbf{s}_{\kk,g_{xy(x^2-y^2)}} \rangle &= \mathbf{B}_{\kk,g_{xy(x^2-y^2)}} \times \langle \mathbf{s}_{\kk,s} \rangle +\mathbf{B}_{\kk,d_{xy}}\times\langle \mathbf{s}_{\kk,d_{x^2-y^2}} \rangle +
	\mathbf{B}_{\kk,d_{x^2-y^2}} \times \langle \mathbf{s}_{\kk,d_{xy}} \rangle+
	\mathbf{B}_{\kk,s}\times\langle \mathbf{s}_{\kk,g_{xy(x^2-y^2)}} \rangle.
	\end{align}
	
	For the $z$-component of the pseudomagnetic field $B_{\kk}^z = \xi_{\kk + \Aq} + \xi_{\kk - \Aq} \equiv \xi_{\kk,\Aq}^s + \xi_{\kk,\Aq}^{d_{x^2 -y^2}} + \xi_{\kk,\Aq}^{d_{xy}} + \xi_{\kk,\Aq}^{g_{xy(x^2 -y^2)}}$ one can integrate out each symmetry. For the chosen dispersion they have the form
	\begin{align}\xi_{\kk,\Aq,s} &= -2t(\cos(A_x) + \cos(A_y))(\cos(k_x) + \cos(k_y)) - 2\mu, \nonumber\\
	\xi_{\kk,\mathbf{A},d_{x^2-y^2}}&= -2t(\cos(A_x) - \cos(A_y))(\cos(k_x) - \cos(k_y)), \nonumber\\
	\xi_{\kk,\Aq,d_{xy}} &= 0 , \nonumber\\
	\xi_{\kk,\Aq,g_{xy(x^2-y^2)}} &= 0.
	\end{align}
	In our model the  occurring order parameters only have either $s$- or $d_{x^2-y^2}$-wave symmetry, which simplifies equations \eqref{eq:eom_full_supp} into 
	\begin{align}
    \frac{d}{dt}\langle \mathbf{s}_{\kk,s} \rangle &= \mathbf{B}_{\kk,s} \times \langle \mathbf{s}_{\kk,s} \rangle + \mathbf{B}_{\kk,d_{x^2-y^2}}\times\langle \mathbf{s}_{\kk,d_{x^2-y^2}} \rangle,
\nonumber\\
\frac{d}{dt}\langle \mathbf{s}_{\kk,d_{x^2-y^2}} \rangle &= \mathbf{B}_{\kk,d_{x^2-y^2}} \times \langle \mathbf{s}_{\kk,s} \rangle + \mathbf{B}_{\kk,s}\times\langle \mathbf{s}_{\kk,d_{x^2-y^2}} \rangle,
	\end{align}
	as those two representations form a subgroup.
	
	\section{Short-time dynamics in the d-wave ground state}
\begin{figure}[h]	
			\includegraphics[width = 0.9\linewidth]{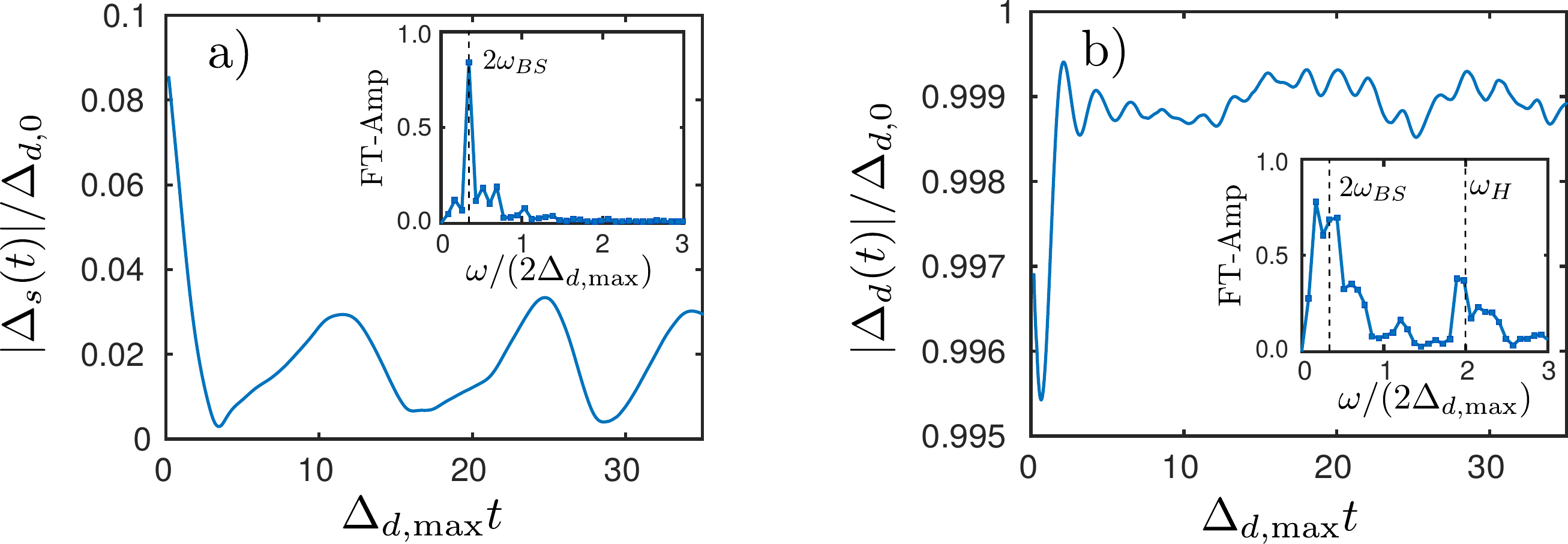}
			\caption{Results of integrating the equations of motions in Eq. (\ref{eq:eom_symc}) for a $d_{x^2-y^2}$-wave ground state at filling $n=0.437$. In a) the short-time dynamics of the induced $s$-wave component of the order parameter and its Fourier transform (inset) are shown. In b) the short-time dynamics of the $d$-wave component and its Fourier transform (inset) are shown.\label{fig:s_in_d}}
		\end{figure}

To verify that the existence of a sub-dominant $s$-wave channel in a $d_{x^2-y^2}$-wave superconductor also gives rise to an additional collective mode in the short-time dynamics we solve Eq. (6) numerically for the $d_{x^2-y^2}$-wave ground state. Due to the nodal gap structure more quasiparticles are excited, which leads stronger damping. However, this also means that the total error bars in the main text are due to both finite accuracy of the Fourier transform and numerical errors due to finite grid size effects.\\
		We consider a $d_{x^2-y^2}$-wave ground state at $n\approx 0.437$ and drive the system out of equilibrium by quenching the $s$-wave order parameter into $\Delta_s = 0.1\Delta_d$. The result in Fig. \ref{fig:s_in_d} shows an induced Higgs mode at $\omega_H = 2|\Delta_{d,\text{max}}(\infty)|$ and also a mode $\omega_{\text{BS}}<\omega_H$. 
		The error bars in the main text are given by the frequency, which is closest to the half width of the peak. Due to the computational error and the stronger damping, the peak of the Fourier transform is broader than in Fig. \ref{fig:d_in_s}. Therefore in addition to the discrete resolution of the Fourier transform the error bars are increased due to computational accuracy.
	
	\section{Linearization of Equations of Motion}
	Here, we perform the linearization of the equations of motion around the equilibrium state at $T=0$. We start with linearizing Eq. (\ref{eq:eom}) and performing a subsequent Fourier transform to obtain
	\begin{align}
	-i\omega\delta\mathbf{s}_\kk = \mathbf{B}^{\text{eq}}_\kk\times\delta\mathbf{s}_\kk +  \delta\mathbf{B}_\kk\times\mathbf{s}^{\text{eq}}_\kk,
	\end{align}
	with the equilibrium pseudomagnetic field $\mathbf{B}_\kk^{\text{eq}} = (2\Delta_{l_0}^\prime\gamma_{\kk,l_0},2\Delta_{l_0}^{\prime\prime}\gamma_{\kk,l_0},2\xi_\kk)^T$ and the equilibrium pseudospin expectation value $\mathbf{s}^{\text{eq}}_\kk = -\mathbf{B}_\kk^{\text{eq}}/(2E_{\kk,l_0})$. We use $l_0$ to denote the groundstate symmetry. After some rearrangement of this expression one obtains a matrix equation
	\begin{align}
	\delta\mathbf{s}_\kk =&\frac{-1}{2E_{\kk,l_0}(\omega^2 - 4E_{\kk,l_0}^2)} \begin{pmatrix}
	-4\xi_\kk^2 & 2i\omega\xi_\kk & -4\Delta_\kk^0\xi_\kk \\
	-2i\omega\xi_\kk & -4E_{\kk,l_0}^2 & -2\Delta_\kk^0 i\omega \\
	-4\Delta_\kk^0\xi_\kk & 2\Delta_\kk^0 i\omega & -4(\Delta_\kk^0)^2
	\end{pmatrix} \begin{pmatrix}
	2\delta\Delta_\kk^\prime \\ 2\delta\Delta_\kk^{\prime\prime} \\
	\delta B_\kk^z
	\end{pmatrix},
	\end{align}
	where $\Delta_\kk^{\text{0}} = \Delta_{l_0}\gamma_{\kk,{l_{0}}}$, which corresponds to the equilibrium order parameter in either $s$-wave or $d_{x^2-y^2}$-wave ground state depending on band filling. Eq. (8) can be then plugged into the definition of the superconducting gap amplitude $	\Delta_l = -\sum_\kk\gamma_{\kk,l} V_l (\delta s_\kk^x -i\delta s_\kk^y)$.
	Consequently, one obtains a linear equation for the real and imaginary part of the gap $\Delta_l$, where the vector $\boldsymbol{\lambda}_l(\omega)$ includes necessary information of the perturbating vector potential in $\delta B^z_\kk(\omega) $.
	\begin{align}
	\begin{pmatrix} \chi_{xx} & \chi_{xy} \\
	\chi_{yx} & \chi_{yy}
	\end{pmatrix} \begin{pmatrix}
	\delta\Delta_l^\prime \\
	\delta\Delta_l^{\prime\prime}
	\end{pmatrix} &= \begin{pmatrix}
	\lambda_l^{\prime} \\
	\lambda_l^{\prime\prime}
	\end{pmatrix} 	\leftrightarrow  \hat{\chi}\delta\mathbf{\Delta}_l = \mathbf{\lambda}_l \quad.
	\end{align}
	The entries of the response function $\hat{\chi}(\omega)$ are given by
	\begin{align}
	\chi_{xx} &= 1+V_l\sum_\kk \frac{4\xi_\kk^2 (\gamma_{\kk,l})^2}{2E_{\kk,l_0}(\omega^2 - 4E_{\kk,l_0}^2)}, \nonumber \\
	\chi_{xy} &= (\chi_{yx})^* = -2V_l\sum_\kk \frac{2\xi_\kk i \omega (\gamma_{\kk,l})^2}{2E_{\kk,l_0}(\omega^2 - 4E_{\kk,l_0}^2)}\nonumber,\\
	\chi_{yy} &= 1+V_l\sum_\kk \frac{4E_{\kk,l_0}^2 (\gamma_{\kk,l})^2}{2E_{\kk,l_0}(\omega^2 - 4E_{\kk,l_0}^2)}
	\end{align}
	for each $l$ respectively. The entries of $\boldsymbol{\lambda}_l(\omega)$ are given by $
	{\lambda_l^\prime = V_l\sum_\kk \frac{2\xi_\kk \gamma_{\kk,l}\Delta_\kk^0\delta B_\kk^z}{2E_{\kk,l_0}(\omega^2 -4E^2_{\kk,l_0})}}$ and $
	{\lambda_l^{\prime\prime} = V_l\sum_\kk \frac{2i\omega \gamma_{\kk,l}\Delta_\kk^0\delta B_\kk^z}{2E_{\kk,l_0}(\omega^2 -4E^2_{\kk,l_0})}}$.
	Collective modes of the system are excited for values of $\omega$ for which the determinant of $\hat{\chi}$ vanishes, i.e., $\det(\hat{\chi}) = 0$ independent of the strength of perturbation.

%\end{document}
\end{document}